\documentclass{aastex}
\usepackage{spr-astr-addons}
\usepackage{verbatim}
\RequirePackage{color}

\newcommand{\emaila}{{Email}:- gchandra.2012@rediffmail.com}
\begin{document}

\title{Photometric Search for variable stars in the field of two Northern open clusters, DOLIDGE 14 and NGC 1960}
\shorttitle{Variables of DOLIDZE~14 and NGC 1960}
\shortauthors{Joshi, G. C.}
\author{Gireesh C. Joshi\altaffilmark{1}}
\affil{$^{1}$Department of Physics, Government Degree College, Kanvaghati (Kotdwar), Pauri-246149}
\emaila

\begin{abstract}
The aim of present work is extract and analyses the light curves of the stars in the field of two clusters, NGC 1960 and DOLIDZE 14. The photometric calibration is performed by comprehensive method of secondary standard transformation and differential photometry using two comparison stars per candidate variable star. The resultant light curves for each potential variable star are displayed and their period analyzed by two different methods. The period and classification of 18 discovered short periodic type variable stars of NGC 1960 are discussed, which consist of four known variable stars and fourteen new variable stars. In the case of DOLIDZE 14, four discovered variables consist of one miscellaneous, one rotational, two $binary$ type variable stars. In the case of NGC~1960, the 12 selected comparison stars appear to be likely candidate for long periodic variability and 4 stars may be standard stars. The variation in brightness of other twenty comparison stars is non-pulsating with an irregular pattern. Membership analysis of variable stars is performed using their distance, kinematic probability and location in $(U-B)$ vs $(B-V)$ TCD. C-M diagrams were constructed to confirm the evolutionary state of the new variable stars.  
\end{abstract}

\keywords{
 Astronomical reduction -- NGC\, 1960, DOLIDZE\, 14, stellar Variability}

\maketitle

\section{Introduction}
\label{int}
An open cluster (OCL) is loosely bounded group of up to a few thousand stars due to the mutual gravitational attraction of cluster members. 
OCLs are host the stars of Pop I. 
The fluctuations in brightness are found for some stars among members of stellar population and such stars are known as variable stars. The stellar variability can arise either due to intrinsic properties (pulsations, eruptions, stellar swelling and shrinking) or due to extrinsic reasons (eclipsed by stellar rotation by another star or planet etc.). 

The variable stars are natural targets of study for any civilization due to their correlation between period and total light output, which allowed them to become the first rung in the astronomical distance ladder \citep{Hippke+2015}. 
Pulsating variables are most important objects due to their periodic expansion and contraction of the surface layers to maintain equilibrium among them. Their census including pulsators and binaries, can provide important clues to stellar evolution and the host star clusters \citep{Luo+2012}. The several classes of pulsating variables are found extensively in the instability strip region of the Hertzsprung-Russell (HR) diagram. Since, pulsating variables have an associated instability strip \citep{du04} above the MS, therefore, an star cluster provides an opportunity to estimate the properties of its stellar variables through its own characteristic parameters. 

Since the detection and magnitude  estimation of the most fainter stars are primary affected by their nearby brighter stars, the knowledge of flux contamination of the stars in the science frame of any cluster is useful for probing the nature of instrumental pseudo-variability. For such a study, a cluster region consisting of bright stars is required and NGC~1960 has been found to be a likely candidate for such a study. Other hand, deep time-series photometry is further bound by exposure time to investigate the variable stars among the fainter stars field of any cluster. For this purpose, DOLIDZE~14 has been found as a possible candidate.

In this background, the time series observations of NGC 1960 and DOLIDZE 14 have been analyzed to search the variable stars within them. The previous parametric studies of both clusters are given in Section 2. The observational details of these clusters are given in Section 3. The methodology of data reduction is discussed in Section 4. The identification procedure of variable stars of DOLIDZE 14 and NGC\,1960 is given in Section 5. Fast-Fourier analysis of variables discuss in Section 6. The mean-proper motions and kinematic membership probabilities for both clusters are described in Section 7. A comparative study of variable stars with parameters of both clusters is discussed in Section 8. A detail description of identified stars of both clusters DOLIDZE~14 and NGC~1960 is given in Sections 9 and 10. The results, discussion and Conclusion are described in Sections 11 \& 12.
\section{Previous studies and antecedents of stellar variability}
\subsection{NGC 1960}
This cluster is located in the Constellation Auriga and has been studied by many authors in the past. The Center coordinates ($\alpha,~ \delta$) for this cluster have been calculated by \cite{sh06} and \cite{ca20} as ($05^{h}:36^{m}:20.8^{s},~ +34^{o}:08^{'}:31^{"}$) and ($05^{h}:36^{m}:20.2^{s},~+34^{o}:08^{'}:06^{"}$), respectively.  Its angular size is computed by \cite{Joshi+2015} and \cite{ca20} as 16 arcmin and 10.3 arcmin, respectively. Previously, this cluster was studied by various research groups [\cite{ba85}, \cite{co17}, \cite{ha05}, \cite{je13}, \cite{jo53}, \cite{kh04} \cite{ni02}, \cite{sa00}, \cite{sh06,sh08}]. 

A complete $UBVRIJHKW_1W_2$ photometric catalogue has been represented by \cite{Joshi+2015}  by complying the PPMXL catalogue with the obtained $UBVRI$ standard photometric magnitude of data collected on date of 30 Nov, 2010 and same data set was further analyzed by JO20 for their absolute/standard photometric analysis. By utilizing catalogues of various data-sets, a comprehensive photometric analysis of this cluster with the long-term variability is shown by them. A total of 76 variable stars of NGC~1960 have been identified by JO20, and their analysis confirmed 72 periodic variables, 59  among them are short period  $(P<1~d)$. They have used absolute phtometry to detect variable stars, in which instrumental errors are surely added due to magnitude transformation. In the case of dataset of J20, there are only three data strings of continuous time series observation with a gape of more than 1 yr and length of each data string is less than 3.5 hours (i.e. $\simeq$ 0.146 d). time gape more than 1 yr leads additional aliases. Mostly, other observational nights have only 1-3 frames with irregular exposure time as well as time interval, which do not seem suitable for the detection of short periodic variable stars. Thus, it is impossible to determine short periodic variable stars.

In the case of this cluster, flux of nearby fainter stars of brighter stars would be contaminated during its deep CCD photometric observations. Such circumstances surely lead an over-estimation in the detection of short periodic variable stars. In the view of above antecedent, author is also motivated to perform time series observations of this cluster with short exposure times of 05, 06 and 10 seconds.     
\subsection{DOLIDZE 14}
In the database of WEBDA, the center coordinates, $(\alpha, ~ \delta)_{J2000}$, of DOLIDZE~14 is ($04^{h}:06^{m}:36.0^{s}, ~+27^{o}:26^{'}:00.0^{"}$) as per work of \cite{alt70}. \cite{jo15c} have been studied the nature of stellar enhancement around the celestial coordinates, $(04^{h}:06^{m}:36.0^{s}, ~+27^{o}:26^{'}:00.0^{"})_{J2000}$ and they shown results as per infrared photometric analysis of stars within DOLIDZE 14. This cluster has shown stellar enhancement in the B-band of USNB1.0, whereas it does not show any stellar enhancement in the infra-red bands \citep{jo15b}. In this connection, they have been estimated the values of $(\alpha,~ \delta)){J2000}$ and ($\mu_{x}, ~\mu_{y}$) as  ($04^{h}:06^{m}:26.7^{s},~+27^{o}:22^{'}:26.7^{"}$) and ($-0.15{\pm}0.34~ mas/yr, ~-7.79{\pm}0.41~mas/yr$), respectively. \cite{jo15c} estimated the values of radius, reddening and distance as $9.6{\pm}0.2~arcmin$, $0.32{\pm}0.02~mag$ and $1.67{\pm}0.14~kpc$ respectively. It is a suitable for analysis of deep CCD-photometic observations due to the system of fainter stars and it is also non-standardized in optical photometry. 

For the name of cluster DOLIDZE~14, \cite{di14} gave the values of center coordinates $(\alpha,~ \delta)_{J2000}$, and proper motions ($\mu_{x}, ~\mu_{y}$) as ($04^{h}:06^{m}:43.0^{s}, ~+27^{o}:32^{'}:34.0^{"}$) and (1.71 $mas/yr$, -0.88 $mas/yr$) respectively. This given center of cluster is separated by 08~arcmin with respect to that of \cite{alt70} and this cluster is denoted by $C~0403+273$ in the database of SIMBAD. As a result, author concludes that cluster DOLIDZE~14 [$(04^{h}:06^{m}:26.7^{s}, ~+ 27^{o}: 22^{'}: 26.7^{"})_{J2000}$] and cluster $C~0403+273$ [$(04^{h}:06^{m}:43.0^{s}, ~+27^{o}:32^{'}:34.0^{"})_{J2000}$] are historically distinct regions and may have cluster properties.

The nature of absolute photometry for time series observation  for any cluster may be understood with respect to secondary standard stars within it and its comparative analysis can be represented using a non-standardized system of stars. As per above antecedents of DOLIDZE~14 [$(04^{h}:06^{m}:26.7^{s}, ~+ 27^{o}: 22^{'}: 26.7^{"})_{J2000}$], it is a likely system of stars for such comparative analysis. 
\section{Data Collection and Extraction}
To detect the short periodic pulsation of stars of target cluster, we need time series observation of the whole night as per availability of target in the telescopic field of view. The time series observations of studied clusters, DOLIDZE 14 and NGC 1960, are carried out using observational facilities of 1.04-m Sampurnand telescope of ARIES, Manora Peak, Nainital. The CCD camera of 1.04-m Sampunanad telescope of ARIES covers $15{\times} 15 ~arcmin^2$ field of view of the target objects. Since, the size of both clusters exceeds the telescopic field of view, therefore, author analyzed the time series observations of the core regions of both clusters. In this connection, the bias and flat frames are also observed for each observational night of studied clusters. The weather conditions (seeing, humidity, wind flow, passing clouds etc.) and declination of target object affect the receiving flux of stars. Thus, the quality of observational data is most important to perform the crucial task of identification of variables. In this connection, the selection procedure of exposure times and observational details of clusters are given below, 

\subsection{Characteristics of observational data of NGC 1960}
%
\begin{figure}
\includegraphics[width=8.3cm]{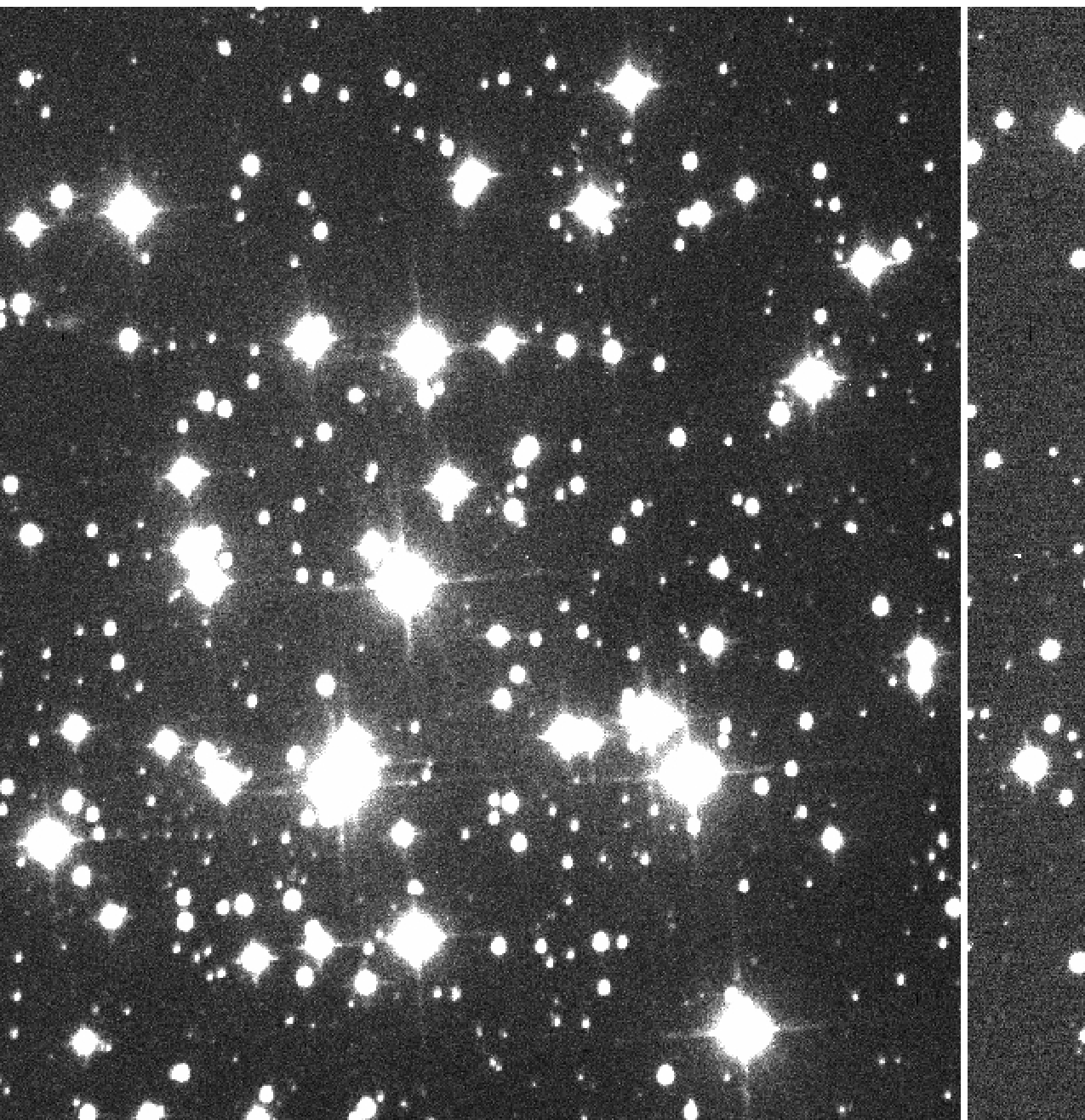}
\caption{In the left and right panels of this figure, the science frames of 60 seconds and 10 seconds are shown for core-region of NGC~1960, which are observed on date 24-01-2012 and 11-12-2013 respectively.}
\label{Fig1}
\end{figure}
%
To identify short periodic pulsations of stars within core region of NGC~1960, time series observations are carried out in V-band during 5 observation nights (2012-2015). It contains ten stars of a visual magnitude brighter than 10 \citep{je13}, one B-type Variable of $9^{th}$ magnitude \citep{de84}, 178 down to magnitude 14 \citep{sa00} and 38 members have infrared excess \citep{sm12}. Thus, there are several bright stars in our telescopic field of view for NGC\,1960. The author found that these bright stars became nearly saturated during an exposure time of 5 seconds. As a result, the value of exposure time of 5 seconds in V-band becomes too high for saturation counts of the brighter stars of NGC \,1960 and leads flux contamination for near by fainter stars of bright stars in the observed science frames using the facility 1.04-m telescope at ARIES, Nainital. Similarly, an exposure time of 1 second is too low value to collection the stellar information for fainter stars of NGC\,1960 below 17 mag in V-band. Environmental influences (seeing, air flow, humidity, passing clouds etc.) and high declination of the target cluster from zenith further reduce the value of stellar magnitude and alter the rate of stellar detection. Therefore, different number of faint stars are detected in different science frames of NGC\,1960. To overcome the detection problem of faint stars, author performed the deep CCD photometric observation of core region of NGC\,1960, with exposure times of 10, 20 and 60 seconds. We need continuous observations of 4-6 hours or more, therefore, the science frames of NGC\,1960  have been captured in the alternating order of low (5 or 6 seconds) and high (10 or 20 or 60 seconds) exposure times during the observation session of night. Thus, the exposure time plays a major role to collect the stellar information. The visual picture of science frames for exposure times of 10 and 60 seconds for NGC~1960 are shown in the right and left panels of Figure~\ref{Fig1}. In these figures, flux contamination of nearby stars of bright stars are found more for science frame with exposure times of 60 seconds than that of 10 seconds. The detail of exposure times and brief description of present data is given in Table 1.
\begin{table*}
\centering
\caption{{\bf The data dated 30/11/2012 and 24/01/2012 for NGC~1960 are common with Joshi \& Tyagi (2015) and Joshi et al. (2020), respectively. In the present work, the author collected an additional data-set of 330 frames over 4 nights in the filter V with $t_{exp}~=~5-60~sec$ (variable).} The observation details of collected data of DOLIDZE\,14 and NGC\,1960 for searching variable stars within them.}
\label{Table1}
\small
{\bf 1. DOLIDZE 14}\\
\begin{tabular}{llllllll}
\hline
S.No. & Date & Observation & Observation  & No. of  & Exposure \\
      &      & Band (Frames) & Time {\&} Mode      & Frames  & Time     \\\hline
  1.  & 13-10-2014 & I & 3.25 hours, Slow & 52 & 150 Sec. \\
  \hline
\end{tabular}\\
{\bf 2. NGC 1960}\\
\begin{tabular}{llllllll}
\hline
S.No. & Date & Observation & Observation  & No. of  & Exposure \\
      &      & Band (Frames) & Time {\&} Mode      & Frames  & Time     \\\hline
1.    & 30-11-2010 & U      & --, Slow & 002   & 300  Sec.  \\
      &            & B      &          & 002   & 300  Sec.  \\
      &            & V      &          & 002   & 200  Sec.  \\
      &            & R      &          & 002   & 200  Sec.  \\         
      &            & U      &          & 002   & 060  Sec.  \\\hline       
2.    & 24-01-2012 & V      & 3.5 hours, Slow & 070   & 60  Sec.  \\\hline
3.    & 11-12-2013 & V (150 frames) & 5.4 hours, Slow & 050   &  05 Sec. \\
      &            &        &                     & 050   &  10 Sec. \\
      &            &        &                     & 050   &  20 Sec. \\ \hline    
4.    & 20-12-2013 & V (080 frames) & 7.6 hours, Slow & 040   &  06 Sec. \\
       &            &        &                     & 040   &  60 Sec. \\      \hline
5.    & 12-01-2015 & V (200 frames) & 7.2 hours, Slow & 100   &  05 Sec. \\
      &            &        &                     & 100   &  20 Sec. \\ \hline  
6.    & 08-02-2015 & V (140 frames) & 5.6 hours, Slow & 140   &  20 Sec. \\     
\hline
\end{tabular}
\end{table*}
After inspecting light curves of studied variable stars and their comparison star in Figures 4-9, the quality of these curves dated 24/01/2012 and 20/12/2013 is too low to identify the nature of stellar variability. An exposure time, 60 seconds, has kept during above observations. Thus, author concludes that observations with exposure times, 05-20 seconds, are suitable to analysis the nature of stellar variability within NGC 1960. In observational sets of 43 nights, JO20 have 1, 1 and 19 observational nights with a exposure time 40, 200 and 60 seconds, respectively. It is noted that data strings for dated 02/11/2011, 03/11/2011 and 24/12/2012 include observation with exposure time 60 seconds. In this background, the classification of variable stars within NGC 1960 by JO20 is seems to be very suspicious.
\begin{figure*}
\centerline{\includegraphics[height=9cm]{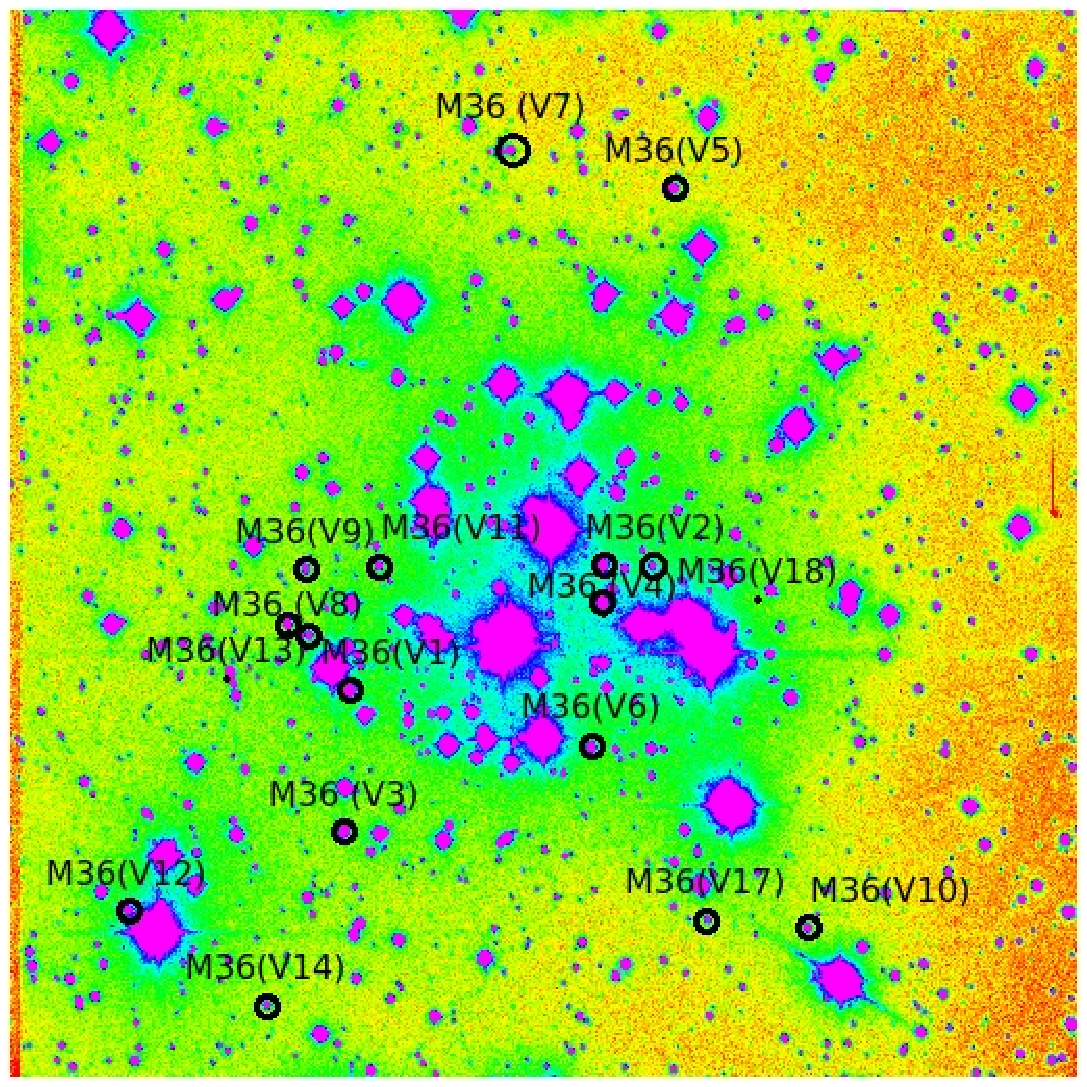}}
\caption{In the left and right panels, the science frames are shown for core-region of NGC~1960 (M36) and DOLIDZE~14 in V-band and I-band as observed on date 30-11-2010 and 13-10-2014 respectively. The exposure times for them were taken to be 200 seconds and 150 seconds, respectively. The present detected variable stars are depicted by open circle in these Finding charts.}
\label{Fig2}
\end{figure*}
%
\subsection{Characteristics of observational data of DOLIDZE 14}
The deep CCD photometric observations are needed for stellar detection in the field of view of DOLIDZE~14 due to its faintness. In this connection, this cluster is observed in I-band on the date 13 Oct, 2014 through 1.04-m Sampurnanand telescope at ARIES, Manora Peak, Nainital. A total of 52 science frames have been captured over a period of 3 hours 15 min. A high value of exposure time (200-300 seconds) is required to identify faint stars of 20 mag or more. It was noted that the positions of the stars slightly shifted during exposure time of 200-300 seconds. Consequently, observations of longer exposure times for open cluster has been avoided. Exposure times of 100-200 seconds are suitable to time series observation of fainter stars, with range of 14-20 mag. Hence, a time series observation of this cluster with an exposure time of 150 seconds has been done by the author. The visual picture of science frame of DOLIDZE~14 is shown in right panel of Figure~\ref{Fig2} and details of its observations is listed in Table~\ref{Table1}.
\section{Methodology of Data reduction}
%
%
%
Bias correction and flat-fielding of observed science frames of NGC \, 1960 and DOLIDZE \, 14 have been performed using bias and flat frames, that were observed in the same observational night of object. Author also utilized bias and flat frames of nearby night for the science frames of NGC \, 1960 due to the lack of these frames in observed data. For this purpose, the `ZEROCOMBINE' and `FLATCOMBINE' tasks of `IRAF' package are used. `COSMICRAYS' task of `IRAF' software are used to remove cosmic rays from the science frames. 
%
%
%
'GEOMAP' and 'GEOTRAN' tasks of IRAF software are used to align the all science frames for analysis. In the astrometry, the pixel coordinates of detected stars are transformed into celestial coordinates ($\alpha_{2000},~\delta_{2000}$) using a linear astrometric solution, derived by matching a set of common stars between present reference catalogue and the 2MASS catalogue with the rms value of about one arcsec in RA and DEC. A total of 29 and 63 common stars have been selected in the observed field of DOLIDZE 14 and NGC 1960 respectively. For this purpose, the visualization of images and access to catalogues has been done by `SKYCAT' tool of ESO\footnote{www.eso.org/sci/observing}. The $CCMAP$ and $CCTRAN$ tasks of $IRAF$ were used for these transformation. 
\subsection{standardization Details for NGC~1960}
To perform consistent photometry from night to night on the aligned images \citep{jo12}, there is need a master list of stars from the science frames of cluster, which have the best seeing and coverage of the observed core region of cluster, NGC~1960. By using prescribed telescope in Section 3.1, the photometric observations of the open star cluster NGC 1960 were obtained on the night of 30 Nov, 2010. The bias and twilight flat frames were acquired during the observational night for the normalization of the CCD pixels. The two Landolt's standard fields $SA95$ and $PG0231+051$ \citep{1992AJ...104..340L} were also observed on the same observational night. A total of ten frames of the cluster with 2 frames each in $U$, $B$, $V$, $R$ and $I$ filters with exposure times of 300, 300, 200, 200 and 60-sec were obtained. All observations were taken in $2$ $\times$ $2$ binning mode to improve the signal-to-noise ratio. The basic steps of image processing such as bias subtraction, flat fielding and cosmic-ray removal were performed through $IRAF$\footnote{ \texttt{Image Reduction and Analysis Facilities (IRAF) is distributed by the National Optical Astronomy Observatories, operated by the Association of Universities for Research in Astronomy Inc., under cooperative agreement with the National Science foundation.}}.  Photometry analysis was done using $DAOPHOT~II$ profile fitting software \citep{1987PASP...99..191S}. To quantify the difference between aperture and profile-fitting magnitudes, an aperture growth curve was constructed by $DAOGROW$ program \citep{1992ASP...25..297S}. The instrumental magnitude was translated into standard magnitude using the following transformation equation:
  \begin{equation}
    m_i = M_i + z_i + c_i \times color + k_i \times X 
  \end{equation}
where $z_i, c_i, m_i, M_i$ and $k_i$ are respectively represent the  zero-point, colour-coefficient, aperture instrumental magnitude and extinction coefficient of different pass-bands. The  $(U-B)$, $(B-V)$, $(V-R)$ and $(R-I)$ colours were used to determine instrumental magnitudes in $U$, $B$, $V$, $R$ and $I$ pass-bands, while $X$ is used for air-mass. The zero-point, colour coefficient and extinction coefficient for $UBVRI$ pass-bands are listed in Table~\ref{Table2}.

\begin{table}
    \caption{The zeropoint, colour-coefficient and extinction-coefficient for different passbands. The colour-coefficients and extinction coefficients listed here.}\label{Table2}
    \medskip
    \tiny
    \begin{center}
      \begin{tabular}{c rrr} \hline
        Filter  & zeropoint($z_i$) & colour  & extinction \\
                &                  & coefficient($c_i$) & coefficient($k_i$) \\\hline
          $U$   & 8.16 $\pm$ 0.01 & -0.05 $\pm$ 0.01 & 0.55 $\pm$ 0.02 ~~~~~~~~~~~~\\
          $B$ &  5.81 $\pm$ 0.02 & -0.01 $\pm$ 0.02 & 0.29 $\pm$ 0.03 ~~~~~~~~~~~~\\
          $V$   & 5.43 $\pm$ 0.01 & -0.08 $\pm$ 0.01 & 0.15 $\pm$ 0.01 ~~~~~~~~~~~~\\
          $R$   & 5.23 $\pm$ 0.01 & -0.09 $\pm$ 0.02 & 0.09 $\pm$ 0.02 ~~~~~~~~~~~~\\
          $I$   & 5.63 $\pm$ 0.02 &  0.01 $\pm$ 0.01 & 0.07 $\pm$ 0.02 ~~~~~~~~~~~~\\\hline
      \end{tabular}\\[5pt]
    \end{center}
\end{table}

\begin{figure*}
\centerline{\includegraphics[height=16cm]{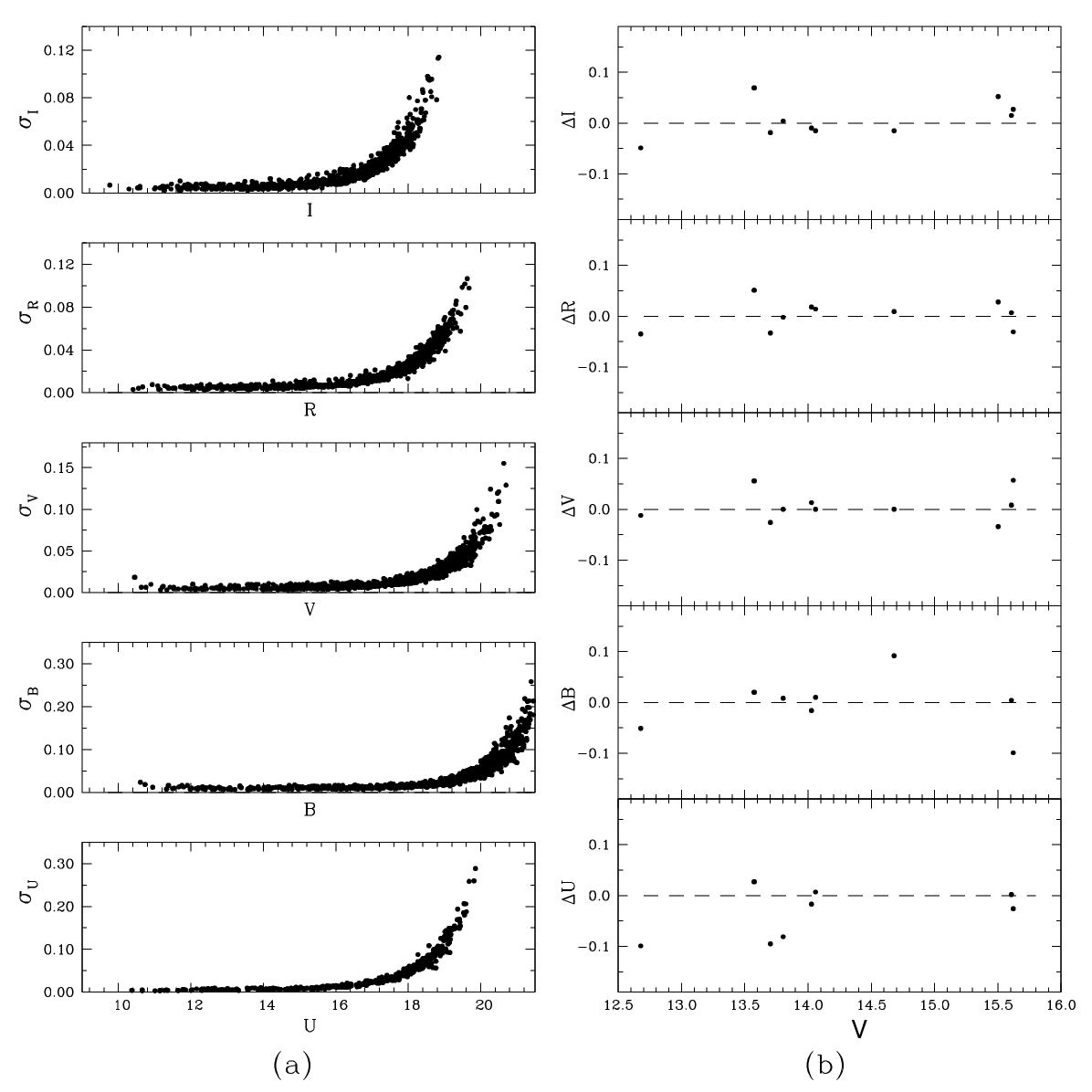}}
\caption{In left panels, we present standard deviation (errors) of stars as a function of brightness. The right panels show difference between our estimated magnitudes with that of the Landolt's magnitude for the standard stars in $UBVRI$ passbands. The black dashed line represents zero shift.}
\label{fig03}
\end{figure*}

In Figure~\ref{fig03}(a), author has shown the variation of standard deviations with brightness of stars in different pass-bands. It is clear from this figure that the errors increases towards the fainter end. The calibrated residuals in magnitude (difference between standard and calibrated magnitude) of standard stars in the Landolt's field are shown in Fig.~\ref{fig03}(b). The standard deviation of the calibration are estimated as 0.083, 0.071, 0.047, 0.030 and 0.049 mag in $U$, $B$, $V$, $R$ and $I$ filters, respectively. The present photometry resulted in a total of 1605 stars within $13'\times13'$ field of the cluster NGC\,1960 in which 447, 1088, 1424, 1583 and 1532 stars were found in $U$, $B$, $V$, $R$ and $I$ bands, respectively. The observed field of NGC~1960 is only central region due to the limited field of view of observation facilities. A total of 1194  stars are found to be common between present photometric data with the \cite{sh06}. In the core region of NGC 1960, 409 additional stars have also been detected in present photometry compare that \cite{sh06}.
%
%
%

%
\begin{table*}
\caption{A complete UBVRI catalogue of the stars in the core field of the cluster NGC\,1960. Columns 2 and 3 are RA and DEC of stars, respectively in epoch J(2000). From column 4 to 13, author gives photometric magnitudes and corresponding errors in $UBVRI$ passbands.}
\label{Table3}
  \small
  \begin{tabular}{rcccccccccccc}\hline
$ID$ & $RA$ & $DEC$ & $U$ & $e_U$ & $B$ & $e_B$ & $V$ & $e_V$ & $R$ & $e_R$ & $I$ & $e_I$   \\\hline    
      1 &  5:36:03.22 & 34:03:37.5 & 10.654 & 0.005 &  10.605 & 0.023 & 10.449 & 0.018 & 10.400  & 0.003 & 10.295 & 0.003 \\
     2  &  5:35:59.29 & 34:10:27.5 & 10.322 & - & - & - & 10.582 & - & 10.543 & - &  10.414 & - \\
     3  &  5:36:03.29 & 34:10:07.9 & 10.457 & - & - & - & 10.601 & - & 10.563 & - & 10.440 & - \\
     4  &  5:36:34.76 & 34:03:55.5 & 10.374 & 0.004 & 10.738 & 0.018 & 10.636 & 0.006 & 10.559 & 0.004 & 10.530 & 0.004 \\
    . &      .      &     .        &      .    &      .    &      .    &      .    &      .    &      .    &      .    &      .    &      .    &      .    \\
     9 &  5:36:08.27 & 34:14:21.2 & 12.547 & 0.005 & 11.900 & 0.016 &  10.900 & 0.010 &  - & -  & 9.762 & 0.007 \\
    10 &  5:36:22.14 & 34:07:13.9 & 11.010 & 0.003 &  11.330 & 0.008 &  11.165 & 0.0032 & 11.082 & 0.007 & 11.019 & 0.004 \\
    11 &  5:36:01.97 & 34:09:17.7 & 11.148 & 0.004 &  11.382 & 0.016 &  11.212 & 0.007 & 11.117 & 0.003 & 10.999 & 0.003 \\
    12 &  5:36:15.80 & 34:14:18.2 & 11.277 & 0.004 &  11.693 & 0.014 &  11.235 & 0.008 & 10.938 & 0.008 & 10.598 & 0.005 \\
    13 &  5:36:11.54 & 34:07:06.5 & 11.145 & 0.003 &  11.496 & 0.009 &  11.340 & 0.003 & 11.274 & 0.007 & 11.210 & 0.005 \\
    14 &  5:35:51.67 & 34:10:32.6 & 11.323 & 0.005 &  11.539 & 0.010 &  11.424 & 0.007 & 11.344 & 0.005 & 11.243 & 0.005 \\
    15 &  5:36:07.75 & 34:09:25.2 & 11.642 & 0.003 &  11.751 & 0.016 &  11.524 & 0.006 & 11.387 & 0.004 &  11.218 & 0.004 \\
    . &      .      &     .        &      .    &      .    &      .    &      .    &      .    &      .    &      .    &      .    &      .    &      .    \\
    . &      .      &     .        &      .    &      .    &      .    &      .    &      .    &      .    &      .    &      .    &      .    &      .    \\
\hline
     \end{tabular}\\[1pt]
\end{table*}

\subsection{Transformation of stellar magnitude for DOLIDZE~14}
It is noted that OCL, DOLDIZE\,14 has not yet been calibrated by any standardized field. In this background, data set of detected stars of its first science frame of our work considered as a reference catalogue for further analysis of stellar variability within it. Other observed science frames of DOLIDZE 14 have been calibrated according to this reference catalogue by using the technique of SSM to reduce atmospheric-effect and estimation-errors of stellar magnitudes during the data collection. For this purpose, a set of common stars is required as per their availability in reference frame and science frames. These common stars are used to find a linear fit between the reference magnitudes and instrumental magnitudes of each frames, assuming that most of the stars have stable magnitude. In this connection, those stars were rejected for linear fitting, which deviate more than 3$\sigma$ limit of deviations of fitting. The resultant linear solution is used to transform instrumental magnitudes of stars of studied clusters into their absolute magnitudes.
\subsection{Secondary standardization method for NGC~1960}
In the case of NGC~1960, to translate the stellar magnitudes (as extracted from data on the remaining nights) into absolute magnitudes, the differential photometry was performed using $UBVRI$ catalogue of secondary stars. For this purpose, author used a linear fit between the standard and instrumental magnitudes on each science frame, assuming that most of the stars are non-variables (these non-variable stars also called stable stars). This procedure is defined as the secondary standardization method [SSM \cite{jo15}]. It is effective to estimate the absolute stellar magnitudes of NGC 1960 through the calibrated magnitudes of its stable stars. The magnitudes of variable stars are rapidly varying compare to other stars and identified variable stars were not utilized for such calibration. In this connection, the master list of stable stars of observed core region of NGC~1960 is prepared by using method as discussed for DOLIDZE~14. 
\section{Identification of variable stars}
The shapes of light curves of a variable star provide valuable information for investigating the nature of stellar variability and underlying physical processes that generate brightness changes. Consequently, the potential variable candidates identify by inspecting of their light curves \citep{Sariya+2014}. If, we find the deviation of absolute magnitudes of star more than 3$\sigma$ limit of mean value of its light curve, then, it would be considered a possible candidate for variable stars. 
The amplitude or period of the pulsations can be related to the luminosity of the pulsating stars and the shape of their light curves can be an indicator of the pulsation mode \citep{wo96}. As a result, pulsating variables are distinguished by duration of their pulsation and the shapes of their light curves \citep{Lata+2014}. For this purpose, the applied procedure for searching variable stars is discussed as below,
\subsection{Need of Differential Photometry for searching variable stars}
The collective information of variation of stellar magnitude over time is known to be light curve of target. The varying sky conditions during the observations generate noise and instrumental errors, which cause the data points in the stellar light curves to be scattered. Thus, stellar light curves carry information of stellar variability as well as irregular variations due to instrument errors, noise and their aliases. Different sky conditions alter the same amount of stellar fluxes for all isolated stars as detected in a science frame of target. The different orders of variation in instrumental magnitudes are obtained for stars as per their different amount of fluxes. Such variation can produce different pattern of pseudo-variability in light curves for stars of different magnitudes. However, the same pattern of pseudo-variability is found for nearby isolated stars, having approximate similar in terms of colour and magnitudes. Hence, such irregular variations have the same pattern for similar isolated stars and can be narrowed down through the differential photometry. If, the comparison stars have been chosen correctly in the case of isolated stars of field of view of any target object, then the difference between their magnitudes should be approximately constant along the night. 
\subsection{Limitation of Differential Photometry in present study}
It is also noted that effect of contamination depends on exposure times as well as stellar distances from the bright stars. Since, the observed field of NGC~1960 is highly contaminated by the presence of bright stars in its core region, therefore, magnitude variation for nearby stars of these bright stars is varied as per physical distance and stellar orientation. Furthermore, exposure time of its science frames is not constant during observation, which further leads to different amount of flux contamination for them. Even for same exposure time, the flux contamination varies with the distance of cluster from Zenith. As a result, difference of instrumental magnitudes of nearby similar comparison stars is not found approximately constant for detected variables of NGC~1960.
Since, DOLIDZE~14 is observed in I-band only, therefore, the comparison stars of its variables are searched in such a way that their I-magnitudes may closer to corresponding variables. Such detected comparison stars were found to be physically distant from their variable. As a result, detected comparison stars are not suitable for variables of DOLIDZE~14. 
\subsection{Secondary standardization Methodology and transformation of stellar magnitudes}
The transformation of apparent magnitude of stars into absolute magnitudes is performed through Secondary Standardization Methodology (SSM). During absolute phototmetry, comparison stars are not required for any variable star and have a major advantage over the differential photometry.
The number of detected stars of any science frame depends on its exposure time. A total of $200{\pm}50$ stars were detected in science frames for DOLIDZE 14, whereas, 1800-3000 stars are found in the science frames of NGC\,1960. By applying SSM, the absolute stellar magnitudes is computed to the instrumental magnitudes of each science frame for NGC\,1960. Other hand, the reference frame of DOLIDZE 14 did not standardized with respect to any standard field stars. The stellar magnitudes of detected stars of each science frame of DOLIDZE 14 are transformed with respect to its reference frame, therefore, SSM method provides the apparent stellar magnitudes for DOLIDZE 14 by considering the uniform variation in stellar magnitudes due to various sky conditions for the entire session of observations.
In these circumferences, the variables stars are detected after visual inspection of light curves as per standard magnitudes via absolute photometry. The possible candidacies of variable stars are determined for stars, having a variation of amplitudes above the 3$\sigma$ limit of its mean in light curves. After visual inspection of the light curves of detected stars of both clusters, author found a total of 4 and 18 possible variable candidates in the observed field of DOLIDZE 14 and NGC 1960, respectively.
\subsection{Limitation of SSM during search of variable stars}
The approximate constant environment parameter (seeing, humidity etc.) and dark night is prerequisites for standardization. The sky conditions change unexpectedly, the transformation coefficients during the process of SSM of each science frame also vary accordingly . As a result, errors of computed stellar magnitudes in absolute photometry can be found due to aliases of different sky conditions with the estimation errors of transformation coefficients. In this connection, light curves of stars show some variation in brightness. These variations are very close to estimated stellar magnitudes with respect to standard and reference catalogue. Such variation can also generate the pseudo stellar variability.
 
These transformed stellar magnitudes are used to generate the light curves of stars. Only those stars are selected as variable stars, that have magnitude variation greater than three times of estimation errors of magnitudes in their light curves as constructed after SSM approach. 
\subsection{Test of stellar variability via SSM and differential photometry}
In the present work, we have not found ideal nearby comparison stars for detected variable stars of both clusters and only differential photometry is not applicable for such case. Since, no available information of influence of the pseudo variability in detected variables, therefore, author also analysis light curves of stable stars of similar order of magnitudes of variables, for tracing the pattern of pseudo-variability among them. 

Since, the transformation of instrumental magnitudes leads additional errors, therefore, the scattering is further increased in data points of their light curves. Such transformation makes a weak information of stellar variability. Thus, it may be possible that a selected stable comparison star has weaker information of stellar variability for any variable star below 3$\sigma$ limit of its light curves. Consequently, we have been avoided the practice of selection of a single stable star for comparing its light curve that of potential variable. For accuracy, author has selected two stable comparison stars for each potentially variable star within their science frames as observed for each cluster. This is done with a view to reducing the difference in magnitudes of selected comparison stars (possibly stable) and their corresponding variable stars.

The pixel distance, differences of stellar magnitudes and colours of identified variables and their comparison stars are listed in the Table~\ref{table4} and Table~\ref{Table5}.

Difference values of pixel coordinated and colours of selected stable stars are consistent with their un-usability for study of variable stars via differential photomety. However, comparative analysis of their light curves becomes a tool to understand the nature of instrumental errors  and to trace the impact of pseudo-variability. In this background, the differential photometry has been performed to confirm the nature of variable stars above the 3$\sigma$ limit of variation of instrumental errors of detected stars within DOLIDZE~14 and NGC~1960.
\subsection{Nature of stellar light curves}
The light curves of potential variable stars and their corresponding stable stars are shown in Figures 4 to 10(A). To distinguish the instrumental variations from stellar light curves, the stellar magnitudes are subtracted from each other and resulting curves are defined as comparative light curves. The author has applied the differential photometry over absolute photometry to construct such comparative light curves. This process is defined as differential-absolute photometry and effectively reduces the effects in comparative light curves due to sky conditions of observational night.

The light curves for each variable and its selected comparison stars (set of three stars) have shown in the different panels of these figures. In this connection, each set of stars have four panels. Top panel shows the light curve of potential variables and middle panels show the light curves of selected comparison stars. The fourth panel of each set have three lines of blue, red and black colour to represent the comparative light curves. The blue and red lines are shown the field subtracted light curves of variable through comparison stars, whereas black line represents the difference of stellar magnitudes of selected comparison stars. A constant spacing of comparative light curves is obtained for stable comparison stars, while the varied spacing of these curves confirms the signature of stellar variability. Since, obtained information of stellar variability changes rapidly with the sky and weather conditions, therefore, we can not find stellar variability of the order of $mmag$ during session of bright moon and observational nights, having fog and high humidity. Consequently, we have selected smoothed light curve to compute the period of identified variable after the visual inspection of individual light curve of each observational night.
\begin{table*}
\centering
\caption{Variable IDs for cluster are listed in first column . V-magnitudes of variable stars of NGC~1960 and I-magnitudes of variable stars of DOLIDZE~14 are given in second column. Values of colour, $(B-V)$, of variable stars of NGC~1960 and colour, $(J-K)$ of variable stars of DOLIDZE~14 are listed in third column. Fourth, fifth and sixth columns indicate the difference of magnitudes for potential variable and its comparison stars. Seventh, eighth and ninth columns represent delta differences of prescribed colour values.  In case of NGC~1960, V-magnitudes are standard magnitude as reported by Joshi \& Tyagi, 2015.}
\label{Table5}
\small
  \begin{center}
{\bf 1:- DOLIDZE 14}\\
\begin{tabular}{cccccccccc}
\hline \hline
Va. & I-magnitude & Colour (J-K) & $\Delta_{I_{mag}}$ & $\Delta_{I_{mag}}$& $\Delta_{I_{mag}}$  & $\Delta_{(J-K)}$ &$\Delta_{(J-K)}$ &$\Delta_{(J-K)}$ \\
ID. & for Variable V   & for Variable V & V {\&} C1 & V {\&} C2 & C1 {\&} C2 & V {\&} C1 & V {\&} C2 & C1 {\&} C2\\
\hline \hline
$V_1$ &  13.784 {$\pm$} 0.015 & 0.264 {$\pm$} 0.038 & -0.370 & -0.483 & -0.113 & -0.394 & -0.487 & -0.093 \\
$V_2$ &  17.385 {$\pm$} 0.030 & 0.656 {$\pm$} 0.072 &  0.013 & 0.011 & -0.002 & -0.206 & -0.173 & 0.033 \\
$V_3$ &  17.804 {$\pm$} 0.055 & 0.139 {$\pm$} 0.183 &  0.003 & -0.006 & -0.009 & 0.074 & 0.301 & 0.227 \\
$V_4$ &  18.501 {$\pm$} 0.086 & 0.595 {$\pm$} 0.139 &  0.032 & -0.001 & -0.033 & 0.396 & -0.133 & -0.529 \\ \hline   
       \end{tabular}\\[5pt]
       {\bf 2:- NGC 1960}\\
    \begin{tabular}{ccccccccc}\hline \hline 
Va. & V-magnitude & Colour (B-V) & $\Delta_{V_{mag}}$ & $\Delta_{V_{mag}}$& $\Delta_{V_{mag}}$  & $\Delta_{(B-V)}$ &$\Delta_{(B-V)}$ &$\Delta_{(B-V)}$ \\
ID. & for Variable V   & for Variable V & V {\&} C1 & V {\&} C2 & C1 {\&} C2 & V {\&} C1 & V {\&} C2 & C1 {\&} C2\\
\hline \hline
$V_{1}$  & 14.007 {$\pm$} 0.004 & 0.511 {$\pm$} 0.005 & -0.032 & -0.017 & -0.015 & -0.164 & -0.078 & 0.086 \\
$V_{2}$  & 14.020 {$\pm$} 0.004 & 0.520 {$\pm$} 0.007 &  0.029 &  0.044 & -0.015 & -0.415 & -0.152 &  0.263 \\
$V_{3}$  & 14.127 {$\pm$} 0.005 & 0.991 {$\pm$} 0.007 & -0.040 & -0.011 & -0.029 &  0.418 & 0.438 & 0.020 \\
$V_{4}$  & 14.215 {$\pm$} 0.008 & 0.569 {$\pm$} 0.008 & -0.036 &  0.019 & -0.022 & -0.055 & -0.145 & -0.123 \\
$V_{5}$  & 14.674 {$\pm$} 0.006 & 0.674 {$\pm$} 0.007 &  0.002 &  0.005 & -0.003 & -0.013 &  0.025 &  0.038 \\
$V_{6}$  & 15.060 {$\pm$} 0.004 & 0.743 {$\pm$} 0.006 &  0.001 &  0.007 & -0.007 & -0.088 & -0.269 & -0.181 \\
$V_{7}$  & 15.155 {$\pm$} 0.009 & 1.522 {$\pm$} 0.014 & -0.001 &  0.004 & -0.005 &  0.760 &  0.509 & -0.251 \\
$V_{8}$  & 15.345 {$\pm$} 0.005 & 0.766 {$\pm$} 0.005 & -0.013 & -0.008 & -0.005 & -0.920 &  0.072 &  0.992 \\
$V_{9}$  & 15.497 {$\pm$} 0.004 & 0.938 {$\pm$} 0.005 & -0.042 &  0.011 & -0.053 & -0.164 & -1.216 & -1.052 \\
$V_{10}$ & 15.592 {$\pm$} 0.004 & 0.778 {$\pm$} 0.005 &  0.005 &  0.022 & -0.017 & -0.064 & -0.027 &  0.037 \\
$V_{11}$ & 15.668 {$\pm$} 0.005 & 0.966 {$\pm$} 0.007 & -0.034 &  0.011 & -0.045 &  0.326 & -0.103 & -0.429 \\
$V_{12}$ & 15.711 {$\pm$} 0.007 & 0.946 {$\pm$} 0.009 &  0.089 &  0.091 &  0.012 &  0.135 & -0.002 & -0.123 \\
$V_{13}$ & 15.769 {$\pm$} 0.004 & 0.820 {$\pm$} 0.006 & -0.001 &  0.031 &  0.009 & -0.364 & -0.032 &  0.373 \\
$V_{14}$ & 15.969 {$\pm$} 0.004 & 0.944 {$\pm$} 0.007 & -0.017 &  0.001 & -0.018 & -0.669 & -0.061 &  0.608 \\
$V_{15}$ & 16.197 {$\pm$} 0.005 & 0.809 {$\pm$} 0.008 & -0.002 &  0.016 & -0.018 & -0.387 & -0.217 &  0.170 \\
$V_{16}$ & 16.279 {$\pm$} 0.006 & 0.984 {$\pm$} 0.008 &  0.022 &  0.041 & -0.019 & -0.001 & -0.709 & -0.708 \\
$V_{17}$ & 16.369 {$\pm$} 0.007 & 1.070 {$\pm$} 0.011 & -0.007 &  0.022 & -0.029 &  0.285 & -0.840 & -1.125 \\
$V_{18}$ & 16.673 {$\pm$} 0.007 & 1.112 {$\pm$} 0.009 & -0.007 &  0.006 & -0.013 &  0.242 &  0.224 & -0.018 \\ \hline \hline  
       \end{tabular}\\[5pt]  
  \end{center} 
\end{table*}
\begin{table*}
  \caption{The first column shows variabe ID of variables within studied clusters. The second and third columns represent $RA$ and $DEC$ respectively. The values of period of detected variable stars are estimated through the PERIOD04 and PerSea Software as listed in fourth and seventh columns respectively.}\label{table3}
  \medskip
  \small
  \begin{center}
  {\bf 1:- DOLIDZE 14}\\
    \begin{tabular}{ccccccc}\hline 
  \hline
Variable & RA & DEC & Period (days) & Amplitude & Power & Period (days)\\
  ID     & ($J2000$)   &  ($J2000$)   & PERIOD04      & (mmag)    & [PERIOD04] & PerSea\\
  \hline
\hline%
$V_{1}$ & $4:07:12.11$ & $27:15:17.7$ & $0.0599{\pm}0.0067$ & $270$ & 04.356 & $0.0606{\pm}0.0068$\\
$V_{2}$ & $4:06:56.99$ & $27:13:13.4$ & $0.0939{\pm}0.0195$ & $474$ & 03.051 & $0.0714{\pm}0.0074$\\
$V_{3}$ & $4:07:02.32$ & $27:24:27.8$ & $0.1349{\pm}0.0359?$ & $311$& 12.169 & $0.1428{\pm}0.0173?$\\
$V_{4}$ & $4:06:21.87$ & $27:26:01.5$ & $0.0674{\pm}0.0085$ & $327$ & 07.793 & $0.0714{\pm}0.0062$\\ \hline   
       \end{tabular}\\[5pt]
       {\bf 2:- NGC 1960}\\
    \begin{tabular}{ccccccc}\hline 
  \hline
Variable & RA & DEC & Period (days) & Amplitude & Power & Period (days)\\
  ID     &  ($J2000$)  &  ($J2000$)   & PERIOD04      & (mmag)    & [PERIOD04] & PerSea\\
  \hline
\hline%
$V_{1}$   & $05:36:25.11$ & $34:06:10.2$ & $0.3057{\pm}0.0815$ & $102$ & 66.948  & $0.4000{\pm}0.1055$\\
$V_{2}$   & $05:36:17.85$ & $34:09:14.8$ & $0.2246{\pm}0.0599$ & $165$ & 56.576  &  $0.6250{\pm}0.0274$ \\
$V_{3}$   & $05:36:33.33$ & $34:06:05.4$ & $0.3598{\pm}0.0001$ & $086$ & 99.261  & $0.4000{\pm}0.2086$\\
$V_{4}$   & $05:36:20.05$ & $34:09:14.6$ & $0.3115{\pm}0.1168$ & $062$ & 35.608  & $0.2857{\pm}0.0706$ \\
$V_{5}$   & $05:35:55.79$ & $34:10:07.6$ & $0.3182{\pm}0.0848$ & $248$ & 34.752  & $0.2857{\pm}0.0454$ \\ &  &  &  & & & ($2.4949:^{*2}$) \\
$V_{6}$   & $05:36:28.54$ & $34:09:05.8$ & $0.1528{\pm}0.0194$ & $069$ & 24.174  & $0.1538{\pm}0.0369$ \\
$V_{7}$   & $05:35:53.49$ & $34:08:09.6$ & $0.1747{\pm}0.0254$ & $065$ & 23.504  & $0.1695{\pm}0.0154$ \\
$V_{8}$   & $05:36:21.20$ & $34:05:25.4$ & $0.2864{\pm}0.0007$ & $073$ & 65.225  & $0.2857{\pm}0.0947$ \\
$V_{9}$   & $05:36:17.96$ & $34:05:38.7$ & $0.2667{\pm}0.0006$ & $095$ & 48.289  & $0.4629{\pm}0.0.0399$ \\
$V_{10}$  & $05:36:39.17$ & $34:11:42.8$ & $0.1886{\pm}0.0003$ & $074$ & 31.753  & $0.2404{\pm}0.0226$ \\
$V_{11}$  & $05:36:17.84$ & $34:06:31.8$ & $1.1053{\pm}0.0007$ & $124$ & 185.267 & $1.098{\pm}0.1033$ \\
$V_{12}$  & $05:36:37.89$ & $34:03:27.5$ & $0.8538{\pm}0.0004$ & $084$ & 93.789  & $0.6667{\pm}1.0505$ \\
$V_{13}$  & $05:36:21.85$ & $34:05:40.9$ & $0.3057{\pm}0.0815$ & $076$ & 35.455  & $0.2857{\pm}0.0762$ \\
$V_{14}$  & $05:36:43.52$ & $34:05:08.8$ & $0.2665{\pm}0.0711$ & $077$ & 13.036  & $0.2222{\pm}0.0775$ \\
$V_{15}$  & $05:36:21.81$ & $34:05:34.1$ & $0.3039{\pm}0.0810$ & $072$ & 23.669  & $0.2041{\pm}0.0.0105$ \\
$V_{16}$  & $05:35:44.69$ & $34:03:03.4$ & $0.3005{\pm}0.0801$ & $087$ & 22.739  & $0.2941{\pm}0.0171$ \\
$V_{17}$  & $05:36:38.67$ & $34:10:29.8$ & $----$ & $----$ & $---$& $----$ \\
$V_{18}$  & $05:36:17.92$ & $34:09:51.1$ & $----$ & $----$ & $---$& $----$ \\ 
\\ \hline   
       \end{tabular}\\[5pt]  
  \end{center} 
\end{table*} 
The period values are determined by two different methodology as statistical and ANOVA analysis. The period values are not agreeing with each other in 6 cases ($V_{2}$, $V_{3}$, $V_{9}$, $V_{10}$, $V_{12}$, $V_{15}$) among variable stars within Field of View of NGC 1960. In the case of $V_{2}$, $V_{10}$ and $V_{12}$ of NGC 1960, both values are seems to be considerable for periodic analysis. It is impossible to determine the true period for variable star $V_{3}$ of NGC 1960 as the nightly data strings are about the same computed values.
\subsection{Comparative Analysis of SSM with essential conditions of Differential Photometry}
The major characteristics of comparative light curves of present comparison stars of variable stars of NGC~1960 are obtained as below,\\
{\bf (1)-} The shifted and varied magnitude differences are found in comparative light curves of comparison stars of variable stars $V_{1}$, $V_{2}$ and $V_{12}$  .\\
{\bf (2)-} A constant value of magnitude difference is found in comparative light curves of comparison stars of variable stars $V_{6}$, $V_{8}$ and $V_{11}$ during observations of individual night. However, shifting of magnitude difference is altered night to night observations.\\
{\bf (3)-} A constant value of magnitude differences is found in comparative light curves of comparison stars of variable $V_{3}$, $V_{4}$, $V_{9}$, $V_{14}$ and $V_{17}$ during the observations on date 24-01-2012, 11-12-2013 and 20-12-2013. Similarly, a shifted and constant value of magnitude differences is obtained in light curves of these stars during the observations on date 12-01-2015 and 08-02-2015.\\
{\bf (4)-} A constant value of magnitude differences is found in comparative light curves of comparison stars of variable $V_{5}$, $V_{9}$, $V_{10}$, $V_{13}$, $V_{15}$, $V_{16}$ and $V_{18}$ during the observations.\\ \\
Thus, detected variable stars of NGC~1960 are listed in four different groups as per comparative light curves of their comparison stars. After deep investigation of Table~\ref{table4} and Table~\ref{Table5}, we did not find any criteria of geometric distribution of comparison stars and their colour-difference values for separating variable stars of NGC~1960 into these obtained groups. It indicates that there are no need of comparison stars for any variable star after implication of SSM approach. In nutshell, the present SSM approach is seems to be reliable for evaluating the stellar variable nature within studied clusters.
\section{Fourier Transform of variables and their Pulsations}
Aliases frequencies occur in the light curves of stars due to the interaction of pulsation of variables and the noise or instrumental errors. Such summation of noise and pulsation signal of variable is removed through of comparison star during differentiate photometry and is an effective method for reducing the uncertainty of detected pulsation signal in the scattered data points of light curves of variables. After confirming the pulsation signal of stars, we need a periodogram to estimate the spectral density of a signal during the pulsation signal processing. Now days, the periodogram are computed from the stellar light curves through the implemented of algorithms such as  Lomb–Scargle folding \citep{Lomb+1976,Scargle+1982}, Box-fitting Least Squares or "BLS" \citep{ko02} and Plavchan \citep{pl08}. Standard and advanced Fourier transform techniques are useful in the analysis of astrophysical time series of very long duration \citep{Ransom+2002} due to their better computing ability. The Lomb-Scargle algorithm is a variation of the Discrete Fourier Transform (DFT), which  decomposes a time series into a linear combination of sinusoidal functions\footnote{exoplanetarchive.ipac.caltech.edu/docs/pgram}. This algorithm is implemented by us to detect pulsation of variables and to construct the Fourier-Discrete- periodogram (FDP). In this connection, the 'PERIOD-04'\footnote{www.univie.ac.at/tops/Period04} and 'PerSea'\footnote{www.home.umk.pl/~gmac/SAVS/soft.html} software are used to estimate the period of new identified variable stars. 'Period04' is dedicated to the statistical analysis of large astronomical time series with gaps and offers tools to extract the individual frequencies from the multi-periodic content. Other hand, 'PerSea' is based on the analysis of variance (ANOVA) algorithm. In Table 5, author has listed the resultant estimated period of variables through the both software. The phase-folded diagrams of detected regular variables are constructed using the values of pulsation period as per 'Period04'. The phase-folded light curves of variables of DOLIDZE 14 are shown in the Figure 10(B), whereas these curves of variables of NGC\,1960  are shown in the Figure 11(A). In these diagrams, the phase values of any variable at time $t$ is defined as decimal part of $(t-JD)/P$, where $JD$ and $P$ represent the Initial Julian Date and Period of the variables. In this connection, the $JD$ values for NGC\,1960 and DOLIDZE\,14 are 2455951.11037 and 2456943.35851, respectively. 
\begin{table*}
  \caption{Variable IDs for cluster are listed in first column. Second, third and fourth columns indicate the {\bf separation pixel-distances} for potential short periodic variable stars and its comparison stars.}\label{table4}
  \medskip
  \small
  \begin{center}
  {\bf 1:- DOLIDZE 14}\\
    \begin{tabular}{ccccccc}\hline 
  \hline
 Variable & $\Delta~D$ & $\Delta~D$ & $\Delta~D$ & SNR & Variable Type\\
 ID  & (in pixel) & (in pixel) & (in pixel) &  &  \\   \hline \hline
 $V_{1}$  & 830.818 & 540.589 & 294.191 & 18.00 & $Miscellaneous$ \\
 $V_{2}$  & 235.701 & 808.347 & 682.126 & 15.80 & $Rotational$ \\
 $V_{3}$  & 466.591 & 657.623 & 281.851 & 05.65 & $Miscellaneous$ \\
 $V_{4}$  & 920.656 & 954.320 & 716.691 & 03.80 & White Dwarfs \\
  \hline   
       \end{tabular}\\[5pt]
       {\bf 2:- NGC 1960}\\
    \begin{tabular}{ccccccc}\hline 
  \hline
 Variable & $\Delta~D$ & $\Delta~D$ & $\Delta~D$ & SNR & Variable Type\\
 ID  & (in pixel) & (in pixel) & (in pixel) &  &  \\   \hline \hline
 $V_{1}$  & 234.715 & 373.093 & 262.214 & 25.50 & $Miscellaneous$ \\
 $V_{2}$  & 189.041 & 509.045 & 596.828 & 41.25 & $\gamma-Dor$ \\
 $V_{3}$  & 448.043 & 145.154 & 334.378 & 17.20 & $Miscellaneous$ \\
 $V_{4}$  & 231.421 & 389.344 & 202.745 & 07.75 & $\gamma-Dor$ (JO20)\\
 $V_{5}$  & 483.260 & 158.671 & 444.375 & 41.33 & EB?\\
 $V_{6}$  & 339.534 & 659.349 & 792.740 & 17.25 & $Ellipsoidal$ \\
 $V_{7}$  & 648.068 & 916.607 & 860.138 & 07.22 & $Ellipsoidal$ \\
 $V_{8}$  & 145.462 & 373.379 & 261.138 & 14.60 & $Rotational$\\
 $V_{9}$  & 643.228 & 392.418 & 753.954 & 23.75 & $RRC$ \\
 $V_{10}$ & 128.526 & 925.765 & 814.031 & 18.50 & $LADS$ \\
 $V_{11}$ & 132.993 & 483.626 & 614.331 & 24.80 & $EB$ \\
 $V_{12}$ & 445.491 & 756.433 & 312.356 & 12.00 & $Rotational$ \\
 $V_{13}$ & 554.006 & 448.464 & 267.522 & 19.00 & $Miscellaneous$ \\
 $V_{14}$ & 635.751 & 813.743 & 298.683 & 19.25 & $Miscellaneous$ \\
 $V_{15}$ & 272.533 & 569.389 & 297.607 & 14.40 & $LADS$ \\
 $V_{16}$ & 763.334 & 599.666 & 346.557 & 14.50 & $Miscellaneous$ \\
 $V_{17}$ & 466.030 & 782.811 & 320.374 & $----$ & Irregular\\
 $V_{18}$ & 291.715 & 398.561 & 640.928 & $----$ & Irregular\\
  \hline   
\end{tabular}\\[5pt]
\end{center} 
\end{table*} 
\subsection{Smoothness of phase diagrams and change in amplitude of pulsation}
There is too much scatter of data points in the original phase diagrams to probe and to shape the nature of stellar variability. Such scattered data points in the curves are due to instrumental errors and noise, due to which, it is not possible to accurately classify the nature of stellar variability. To overcome this problem, we adopt the average moving procedure to construct these diagrams. In this process, data points of phase are arranged in increasing order from 0 to 1. In this connection, the average values are determined for sets of five data points such as 1-5, 2-6, 3-7 and so on. This process is repeated until the average of last remaining five data points has been calculated. However, a sufficient fraction of the amplitude of light curve also decreases during this process. The resultant phase-folded curves of variables are found to be smoother compare than original diagrams. As a result, it is concluded that amplitude of stellar pulsation decreases with the increment of smoothness of the phase-folded diagram of variables during the moving average procedure. In the Figures 10(B) and 11(A), the phase diagrams of variables are constructed through the resultant data points as per the average moving procedure. The signal-to-noise ratio ($SNR$) is defined as $SNR=\frac{A}{e_A}$, where $A$ and $e_A$ are amplitude of light curve and mean estimation error in stellar magnitude respectively. The $SNR$ values of variable stars are listed in Table 6. 
%
%
%
\section{Mean proper Motion and Membership Analysis}
\subsection{Mean Proper Motion of Core region of NGC\,1960}
A proper motion study of this cluster was done by \cite{me58}, \cite{ch66} \cite{sa00} and \cite{jo20} [JO20 now onward]. Author compared present catalogue with Gaia EDR3 data to that given by \citet{gaia16,gaia21} and found 1579 common stars between them within the studied core region. The distribution of these stars in ${\mu_x}$-${\mu_y}$ plane is shown in the Figure~15. The dark filled circles in the figure are those stars that were used to determine the mean proper-motion of the cluster NGC~1960. These stars were identified after excluding stars located outside 3$\sigma$ deviation from the mean value of proper motion in both $RA$ and $DEC$ directions. After 3$\sigma$ iteration processes, the author found 1405 stars that were used to obtain following mean proper motion in $RA$ and $DEC$ direction of the cluster NGC~1960.
$$
\bar{\mu_{x}}, ~ \bar{\mu_{y}} =0.314\pm0.026~mas/yr,~-2.333\pm0.037~mas/yr
$$
JO20 has been estimated the mean proper motion of whole cluster of NGC~1960 in $RA$ and $DEC$ directions as $-0.143{\pm}0.008~mas~yr^{-1}$ and $-3.395{\pm}0.008$ respectively. Since, JO20 shows a very different selection of cluster members that do not include stars centered around zero pm, therefore their different proper motion values are obvious due to the selection method applied to the stars. Due to the inclusion of field stars, the standard deviation values ($\sigma_{\alpha},~\sigma_{\delta}$) of stellar proper motions of NGC 1960 are found to be 0.996 and 1.402 in RA and DEC respectively. Using these values, the resultant standard deviation $\sigma= \sqrt{\sigma_{\alpha}^2 + \sigma_{\delta}^2}$ is estimated to be 1.734. This standard deviations of the stellar proper motion of cluster NGC 1960 are not compatible with the standard errors of present work as well as the work of JO20. Thus, the different values of proper motions of both studies indicates that the stellar members of this cluster may segregated in inner and outer regions according their proper motion values.     
\subsection{Mean Proper Motion of DOLIDZE\,14}
In the case of DOLIDZE\, 14, a total of 1137 stars are found  within periphery of radius $9.8~arcmin$ in the GAIA database. Author found proper motions for 887 stars among them, which are used to compute the mean proper motion of DOLIDZE\,14. After 3$\sigma$ iteration processes, author found 761 stars from and used to obtain following mean proper motion in $RA$ and $DEC$ direction of the cluster DOLIDZE~14.
$$
\bar{\mu_{x}}, ~\bar{\mu_{y}}=1.050\pm0.083~mas/yr,~-2.254\pm0.093~mas/yr
$$
Due to the inclusion of field stars, the estimated standard deviation values ($\sigma_{\alpha},~\sigma_{\delta}$) of proper motion values of stars of DOLIDZE~14 are 2.288 and 2.601 in RA and DEC respectively. These values give resultant standard deviation $\sigma= \sqrt{\sigma_{\alpha}^2 + \sigma_{\delta}^2}$ as 3.465. These values are not compatible with the standard errors of the stellar proper motion of DOLIDZE~14 and confirms the proper decontamination process due to field stars within cluster region.
\subsection{Kinematic Probabilities}
In the present analysis, those stars are consider as kinematic members for each cluster, which are within 3{$\sigma$ limit of the mean proper motion of studied cluster. The proper motion probability is assigned to be 1 for stars that lie within 3{$\sigma$} limit of mean proper motion for the cluster, whereas the proper motion probability is assigned to be 0 for stars with proper motions outside the 3{$\sigma$} limit \citep{jo22}. {\bf By utilizing extracted stellar proper motion from GAIA EDR3 database}, new kinematic probabilities of stellar members of each cluster are computed as
$P_{pm}= 1-\frac{\sqrt{(\mu_{\alpha}- \bar{\mu_{\alpha}})^2 + (\mu_{\delta}- \bar{\mu_{\delta}})^2}}{3\sigma}$. These values of variable stars of both clusters are listed in sixth column of Table~7.
\subsection{Stellar Parallax and Membership}
The individual stellar distance of each potential variable star is computed using the parallax information extracted from GAIA EDR3 database and are listed in third column of Table~7. These values are compared with the estimated distance for concerned cluster via CMD analysis. Positional membership of cluster is assigned for those variable stars, whom distance is equal to that of its parent cluster and no positional membership is assigned for other variable stars. A comparative description of the kinematic and positional probabilities for each individual variable stars is given in Sections 9 {\&} 10.  
\begin{table*}
  \caption{Prallax and distance values of studied variable stars are listed in second and third columns respectively. The stelar proper motion values of R.A. and DEC deirection are given in fourth and fifth columns respectively. The estimated kinematic probabilities of variables are listed in sixth column. }\label{table8}
  \medskip
  \small
  \begin{center}
  {\bf 1:- DOLIDZE 14}\\
    \begin{tabular}{ccccccc}\hline 
  \hline
 Variable & Parallax & Distance & Proper Motion  & Proper Motion & kinematic & kinematic \\
 ID & (mas) & (kpc) & in RA ($\delta$) & in DEC ($\mu$) &  Pro. (Old) &  Pro. (GAIA) \\  \hline \hline
 $V_{1}$  & $1.2234{\pm}0.0162$ & $0.817{\pm}0.011$ & $ 2.317 {\pm} 0.019$ & $ -7.763 {\pm} 0.012$ & $--$ & 0.456 \\
 $V_{2}$  & $0.7388{\pm}0.0772$ & $1.354{\pm}0.141$ & $ 1.903 {\pm} 0.094$ & $  1.060 {\pm} 0.054$ & $--$ & 0.671 \\
 $V_{3}$  & $0.2453{\pm}0.0709$ & $4.077{\pm}1.178$ &  $ 0.590 {\pm} 0.077$ & $ -0.856 {\pm} 0.046$ & $--$ & 0.858 \\
 $V_{4}$  & $0.9287{\pm}0.1791$ & $1.077{\pm}0.208$ &  $-1.934 {\pm} 0.204$ & $ -3.928 {\pm} 0.116$ & $--$ & 0.671 \\
   \hline   
       \end{tabular}\\[5pt] 
      {\bf 2:- NGC 1960}\\
    \begin{tabular}{ccccccc}\hline 
  \hline
 Variable & Parallax & Distance & Proper Motion  & Proper Motion & kinematic & kinematic \\
 ID & (mas) & (kpc) & in RA ($\delta$) & in DEC ($\mu$) &  Pro. (Old) &  Pro. (GAIA) \\  \hline \hline
 $V_{1}$  & $0.8858{\pm}0.0188$ & $1.129{\pm}0.024$ &  $-0.204 {\pm} 0.023$ & $ -3.406 {\pm} 0.016$ & 0.40 & 0.771 \\
 $V_{2}$  & $0.8430{\pm}0.0180$ & $1.186{\pm}0.025$ &  $-0.011 {\pm} 0.023$ & $ -3.506 {\pm} 0.017$ & 0.35 & 0.766 \\
 $V_{3}$  & $2.9243{\pm}0.0228$ & $0.342{\pm}0.003$ &  $-3.018 {\pm} 0.027$ & $ -4.932 {\pm} 0.019$ & 0.06 & 0.188 \\
 $V_{4}$  & $0.8382{\pm}0.0194$ & $1.193{\pm}0.027$ & $-0.432 {\pm} 0.024$ & $ -3.219 {\pm} 0.017$ & 0.99 & 0.777 \\
 $V_{5}$  & $0.2889{\pm}0.0255$ & $3.461{\pm}0.305$ & $ 0.675 {\pm} 0.028$ & $ -2.292 {\pm} 0.021$ & 0.05 & 0.930 \\
 $V_{6}$  & $0.8543{\pm}0.0254$ & $1.171{\pm}0.035$ & $-0.209 {\pm} 0.034$ & $ -3.440 {\pm} 0.023$ & 0.81 & 0.765 \\
 $V_{7}$  & $0.1985{\pm}0.0275$ & $5.038{\pm}0.698$ & $ 0.113 {\pm} 0.027$ & $ -1.594 {\pm} 0.021$ & 0.00& 0.853 \\
 $V_{8}$  & $0.8206{\pm}0.0350$ & $1.219{\pm}0.052$ & $-0.042 {\pm} 0.040$ & $ -3.525 {\pm} 0.026$ & 0.79 & 0.761 \\
 $V_{9}$  & $0.4966{\pm}0.0857$ & $2.014{\pm}0.348$ & $ 0.010 {\pm} 0.097$ & $ -3.579 {\pm} 0.072$ & 0.91 & 0.753 \\
 $V_{10}$ & $0.8022{\pm}0.0309$ & $1.246{\pm}0.048$ & $-0.244 {\pm} 0.036$ & $ -3.276 {\pm} 0.027$ & 0.04 & 0.789 \\
 $V_{11}$ & $0.8130{\pm}0.0350$ & $1.230{\pm}0.053$ & $-0.401 {\pm} 0.043$ & $ -3.543 {\pm} 0.030$ & 0.94 & 0.730 \\
 $V_{12}$ & $0.9063{\pm}0.0400$ & $1.103{\pm}0.049$ & $-0.059 {\pm} 0.049$ & $ -3.355 {\pm} 0.035$ & 0.00 & 0.791 \\
 $V_{13}$ & $0.4407{\pm}0.0366$ & $2.269{\pm}0.188$ & $ 0.723 {\pm} 0.041$ & $ -3.327 {\pm} 0.029$ & 0.00 & 0.793 \\
 $V_{14}$ & $0.0995{\pm}0.1441$ & $10.050{\pm}14.555$ & $ 0.692 {\pm} 0.159$ & $ -2.518 {\pm} 0.128$ & 0.99 & 0.919 \\
 $V_{15}$ & $0.9037{\pm}0.0496$ & $1.106{\pm}0.061$ & $-0.331 {\pm} 0.055$ & $ -3.593 {\pm} 0.037$ & 0.00 & 0.728 \\
 $V_{16}$ & $1.1574{\pm}0.0478$ & $0.864{\pm}0.036$ & $ 4.521 {\pm} 0.057$ & $ -1.697 {\pm} 0.038$ & 0.93 & 0.182 \\
 $V_{17}$ & $0.9037{\pm}0.0448$ & $1.107{\pm}0.055$ & $-0.178 {\pm} 0.055$ & $ -3.418 {\pm} 0.040$ & 0.52 & 0.771 \\
 $V_{18}$ & $0.8396{\pm}0.0460$ & $1.191{\pm}0.065$ & $-0.246 {\pm} 0.054$ & $ -3.425 {\pm} 0.039$ & 0.95 & 0.764 \\
  \hline   
\end{tabular}\\[5pt]
\end{center} 
\end{table*} 
\section{Comparative study of cluster's parameters with variable stars}
\subsection{Variable Stars and ZAMS of NGC~1960}
By using $(U-B)$ vs $(B-V)$ Two colour Diagram (TCD), \cite{Joshi+2015} has been determined reddening, $E(B-V)=0.23{\pm}0.02 ~mag$. JO20 has also computed the reddening, $0.24{\pm}0.02~mag$, which is close to the previous one. In the present work, the location of variable stars of NGC~1960 is shown in $(U-B)$ vs $(B-V)$ TCD along-with Zero-Age-Main-Sequence (ZAMS) as shown in Figure~\ref{fig15}. A ZAMS star has its minimum radius, its maximum mass (for single star evolution), its bluest colour (or hottest effective temperature), and its central core possesses its peak $H/He$ \citep{ha98}. After deep inspection of TCD and GAIA database for cluster, author has been drawn following facts:\\
{\bf (1)-} Variable stars $V_1$, $V_2$, $V_3$. $V_4$ are found to be ZAMS stars and have close brightness to each other in V-band. The locations of $V_1$, $V_2$ and $V_4$ at the near of bump in $(U-B)$ vs $(B-V)$ TCD, whereas the location of $V_3$ is found to be far away these stars. This facts leads different class of $V_3$.\\
{\bf (2)-} Variable stars $V_8$, $V_{12}$, $V_{13}$, $V_{14}$ and $V_{16}$ are belong to ZAMS of cluster. Variable stars $V_8$ and $V_{12}$ are member of cluster ~1960 and may poses the character of same class of variable stars. Variable stars $V_{13}$, $V_{14}$ and $V_{16}$ are field stars leading to the fact that field stars are also located in just main sequence of cluster.\\ 
{\bf (3)-} Variable stars $V_6$, $V_{10}$, $V_{11}$, $V_{15}$, $V_{17}$ and $V_{18}$ do not belong to main sequence of cluster and have a confirm membership of cluster.\\
{\bf (4)-} Variable stars $V_{5}$, $V_{7}$, and $V_{9}$ neither belong to main sequence of cluster nor are members of cluster.
\subsection{CMDs analysis for NGC 1960}
The log-age of this cluster has been reported by \cite{kh04}, \cite{Joshi+2015} and JO20 as $7.62$ (yr), $7.35$ (yr) and $7.44$ (yr), respectively. \cite{Joshi+2015} have shown the most probable members (MPMs) in $(B-V)$ vs $V$ Colour Magnitude Diagram (CMD) by comprehensive analysis of photometric, kinematic and spatial probabilistic criteria. These MPMs are found along with MS of NGC\,1960 and is also well aligned with a well fitted theoretical isochrone as depicted in upper panels of Figure~\ref{fig16}. The pink line of each CMD represents the well fitted theoretical isochrone as computed by \cite{Joshi+2015}. In the $(J-H)$ vs $H$ CMD, author did not find stellar alignment  by MPMs along with theoretical isochrone for H-magnitude range of 12.5-13.2 $mag$ as shown in Figure \ref{fig16}(B). Thus, an improvement is needed in estimation of distance-modulus. Author overplot Marigo's theoretical isochrones on the CMDs by varying the distance modulus and age simultaneously by keeping reddening $E(B-V)=0.23$ mag. From the best visual isochrone fit to the varying age and distance combinations, author obtained a distance modulus $(V-M_V)=11.05{\pm}0.0.30$~mag and $log(Age)=7.35{\pm}0.05$~(yr) for cluster NGC\,1960. Employing the correction for the reddening and assuming a normal reddening law, this corresponds to a true distance modulus $(m-M)_0 = 10.34{\pm}0.30$ mag or a distance of $1.169{\pm}0.173~kpc$ for the cluster. {\bf This distance is close to that of $1.17{\pm}0.06~kpc$} as estimated by JO20 via a mean parallax, $0.86{\pm}0.05~mas$, for probable cluster members of NGC 1960 (For this purpose, parallax values were extracted by JO20 from GAIA database).}  The computed value of distance modulus is used to identify the true member of cluster according to retrieved value of distance/parallax value of individual star. 

The best fitted theoretical isochrones in each CMD is represented by black line and identified variable stars are depicted by red dots. Variable stars $V_{3}$ and $V_{7}$ show colour excess in near-infrared bands, however these stars also found along-with the theoretical isochrone in $W_{2} - W_{1}~ vs~ W_{1}$ CMD as depicted in Figure \ref{fig16}(C).
\subsection{Range of Instability strip in the case of NGC~1960}
The instability strip is a narrow, almost vertical region in HR diagram, which includes many different types of variable stars. Most stars more massive than Sun enter the instability and become variable at least once after leaving the main sequence (MS) \footnote{astronomy.swin.edu.au/cosmos/I/Instability Strips}. This strip intersect the MS in the region of A and F stars (masses 1-2 $M_{\odot}$) of studied clusters and extends to G and early K bright super-giants. The common area of instability strip and MS of OCL NGC\,1960 seems to be important region (includes A and F stars) for understanding the cluster dynamics through the stellar variability and vice-versa. To determine the location of the A/F type stars in the colour-magnitude diagram, distance modulus, $(V-M_V)~=~11.05{\pm}0.30~mag$, was applied to obtain the V-magnitudes. In this connection, the upper and lower limit of region, have A and F stars, are found to be $13.17{\pm}0.30 ~mag$ of $V-band$ and $16.61{\pm}0.30~ mag$ of $V-band$, respectively and prescribed intercepted region also least affected by the brighter stars and their neighbourhood. Consequently, author is carried out time series analysis for finding stellar variability within this magnitude-range. A total of eighteen variables of NGC\,1960 have been identified in this magnitude-range. In the present case of NGC~1960, the author did not perform any analysis to search variable stars in near the turn-off region and region of faint stars. 
\subsection{CMDs analysis for DOLIDZE~14}
In the case of DOLIDZE 14, there is no saturated brighter star of unresolved center. As a result, author did not need employ any selection criteria of stellar magnitudes to identify variable stars for this cluster. The red dots of lower panels of Figure~\ref{fig16} represent identified variable stars within the core region of DOLIDZE~14. The distance modulus and distance for cluster DOLIDZE~14 have been estimated by \cite{jo15c} as $11.12{\pm}0.18~mag$ and $1.67{\pm}0.14~kpc$ respectively. In this connection, the best fitted isochrone is depicted by black solid line in Figure \ref{fig16}, whereas pink and blue lines represent lower and upper limits for apparent distance modulus. The distance of identified variable stars of this cluster are listed in table \ref{table8}. Variable star $V_2$ is close to the cluster periphery, whereas variable star $V_4$ appears to be confirmed member of cluster. Other two variable stars $V_{1}$ and $V_{3}$ do not belong to cluster as per known parameters.  
\section{Detected Variables in DOLIDZE 14}
The time series observation of this cluster have been taken in Near-Infrared region using I-band with effective wavelength 8000 $\AA$. In the present analysis, a total of 4 variable stars are found in the field of view of DOLIDZE~14. 
\subsection{Miscellaneous type variable star}
Some periodic stars can not be classified into any particular class of variable stars as per their estimated parameters and phased light curves. 
\subsubsection{Star IDs 004 ($V_1$), 005 and 006 of DOLIDZE~14}
As par 'PERIOD04' and 'PerSea' code analysis, the period values of $V_1$ are found to be $0.0599{\pm}0.0067~d$ and $0.0606{\pm}0.0068$ respectively. Its individual distance value ($0.817{\pm}0.011$) is very far from the cluster distance. Its least kinematic probabilistic value also supports that it is not a member of cluster. Thus, it can not be classified as a main sequence star for the studied cluster. Potential variable $V_1$ is located just near the turn off point of CMD with blue colour. As per analysis of properties of this star, it appears to be a miscellaneous type variable star.\\
Light curve of this variable is compared to the light curves of star IDs 005 ($4^{h}:06^{m}:26.85^{s}, ~+27^{0}:18^{'}:24.2^{"}$) and 006 ($4^{h}:06^{m}:43.39^{s}, ~+27^{0}:17^{'}:46.9^{"}$). Light curves of selected comparison stars show no sign of periodic variability. 
\subsubsection{Star IDs 109, 110 ($V_3$) and 111 of DOLIDZE 14}
Variable star $V_3$ poses a single minima in its light curve. This feature leads to its potential candidature as a variable star. The computed periods are $0.1349{\pm}0.0359~d$ and $0.1428{\pm}0.0173~d$ as per codes 'PEROD04' and 'PerSea' respectively. Both values are approximate similar with length of data string of observation session for DOLIDZE~14 and leads to a dubious analysis for assign classification of its stellar variability. Thus, it is quite impossible to derive a period from a single-night light curve that extent for a few hours, with only a single event of s drop in the light curve. Its distance is found to be $4.077{\pm}1.178~kpc$ and leads it is as a non-member of cluster. As a result, variable star $V_{3}$ may be a system of binary stars with unknown periodic eclipse.\\
Light curves of comparison star IDs 109 ($4^{h}:06^{m}:42.69^{s}, ~+27^{0}:20^{'}:30.2^{"}$) and 111 ($4^{h}:06^{m}:45.90^{s}, ~+27^{0}:17^{'}:00.8^{"}$) show no sign of stellar variability as depicted in Figure \ref{fig10}. 
\subsection{Potential Rotational Variable star}
Large spots in surfaces of Rotational variable stars cause a change in apparent brightness. The light curves of such type variable stars are typically very noisy due to evolution of star spots over time. 
\subsubsection{Star IDs 085, 086 and 088 ($V_2$) of DOLIDZE~14}
A value of period for variable star $V_2$ is calculated by the code of 'PERIOD04' and 'PerSea' to be $0.0939{\pm} 0.0195~d$ and $0.0714{\pm}0.0074$, respectively. Its amplitude, distance and kinematic membership for cluster is computed to be $0.088~mag$, $1.354{\pm}0.141~kpc$ and $0.671$ respectively. Its location in $(J-H)~ vs ~ H $ CMD has found towards red colour as depicted in Figure \ref{fig16}. As a result, this star appears to be a rotational type star. As DOLIDZE~14 is an old cluster with an age of about $1.26{\pm}0.08 ~Gyr$, it is rare/ impossible to find any rotational type variable star in the cluster. Hence, this variable star can not be a member of cluster.\\
In Figure \ref{fig10}, the comparative Light curves of star IDs 085 ($4^{h}:07^{m}:02.90^{s}, ~+27^{0}:15^{'}:53.4^{"}$) and 086 ($4^{h}:06^{m}:38.14^{s}, ~+27^{0}:22^{'}:33.0^{"}$) along with variable star $V_2$ has been depicted. Author did not find any sign of periodic variability in the light curves of said comparison stars.  
\subsection{Potential Binary system of stars}
A sudden drop in brightness of binary system finds due to eclipse and transit of its component stars during their orbit along plane of our line of sight. 
\subsubsection{Star IDs 183, 184($V_4$) and 185 of DOLIDZE 14}
Variable star $V_4$ is not found to be a member of of DOLIDZE 14 as per its distance $1.077{\pm}0.208 ~kpc$ and membership probability $0.671$. By using the codes of 'PERIOD04' and 'PerSea', the period of this variable star is computed as $0.0674{\pm}0.0085~d$ and $0.0714{\pm}0.0062~d$ respectively. The orbital periods of binary stars do not get as shorts as $0.1~days$ unless the components are white dwarfs or similar compact objects. Thus, light curve of variable star $V_4$ poses character of white dwarfs as depicted in Figure \ref{fig10}.\\
Light curves of comparison star IDs 183 ($4^{h}:06^{m}:24.45^{s}, ~+27^{0}:14^{'}:24.9^{"}$) and 185 ($4^{h}:07^{m}:02.11^{s}, ~+27^{0}:17^{'}:52.8^{"}$) are shown the more variation in brightness compare with other comparison stars of studied cluster. However, there are no periodic variation in comparative light curve of comparison stars (Star IDs 183 and 185) and have low variation of brightness compare to their comparative light curves with variable star $V_4$.  
\section{Detected Variables in NGC~1960}
In this paper, eighteen variables were detected in the field of view of NGC~1960, four of them are common with the four variable stars of JO20. According to the behavior of the light curves and the period analysis, the identified variable stars were classified. Among the eighteen detected variables of NGC~1960, one as EB, one as planet transit, one as $\gamma-Dor$, two as $\delta~Scuti-\gamma-Dor$,  two as LADS, two as irregular, two as rotational, three as RRC and four as Ellipsoidal type variable stars.
\subsection{$\gamma-Doradus$ variables}
The light curves of regular variable stars are repeating with a constant value of time (i.e. its period). A multiperiodic stars with g-mode pulsation is called a $\gamma-Doradus$.  These are typically young, early F- or late A-type main stars with periods in the range of about $0.3-3 ~d$ and brightness fluctuation $\sim 0.1 ~mag$ \citep{ba11}. 
\subsubsection{Star ID 606 ($V_2$), 616 and 626 of NGC~1960)}
In the case of variable star ID 606, star IDs 616 ($5^{h}:36^{m}:26.45^{s}, ~+34^{0}:07^{'}:39.3^{"}$) and ID 624 ($5^{h}:35^{m}:50.83^{s}, ~+34^{0}:06^{'}:03.6^{"}$) are selected for comparing their light curves with it. JO20 is assigned ID 606 as a cluster member $\delta-Scuti$ variable star with period $0.27632{\pm}0.00008~d$ and is denoted by $V_{35}$ in their analysis. In the present work, its period is computed as $0.2246{\pm}0.0599~d$ and $0.6250{\pm}0.0274~d$ by the codes of 'PERIOD04' and 'PerSea', respectively. Both computed vales are far from the length of data string, and shows a characteristic of multi-periodicity. Since its light curves in Figure~\ref{fig04} shows no indication of period value less than 7.6 hours, therefore its period value appears to be more accurate via Anova analysis. Thus, Star ID 606 is detected as a $\gamma-Doradus$ variable star instead of $\delta-Scuti$. Its kinematic probability makes it as the most likely member of cluster.

The light curves of ID 616 are showing irregular flux variations, whereas the light curves of ID 624 have characteristics of long periodic type variable stars. Almost constant magnitude was found for ID 624 during 7-8 hours of observations.
\subsection{RR Lyre stars}
Over 80 $\%$ of all variables known in globular clusters are RR Lyre stars \citep{cl01}. One d-type RR Lyre variable star in the open cluster NGC~2141 has been reported by \cite{lu15}. RR Lyrae variables do not follow a strict period-luminosity relationship at visual wavelengths, although they do in the infrared K-band \citep{ca04}.
\subsubsection{Star IDs 1431, 1470 ($V_9$) and 1484 of NGC~1960}
The variable star $V_{9}$ is assigned as a rotational variable star, main sequence star and cluster member by JO20 and marked by $V_{48}$ in their analysis. Its period is computed to be $0.2667{\pm}0.0006~d$ and $0.4629{\pm}0.0399~d$ by the codes of 'PERIOD04' and 'PerSea' respectively. These codes clearly provide two distinct values of period. It seems that computed period $0.2667{\pm}0.0006~d$ ( as per 'PERIOD04' code)is the a average length of data strings in 2013 and 2015, whereas period value $0.4629{\pm}0.0399~d$ (as per 'PerSea' code) is cross matched with period $0.47483{\pm}0.00023$ as estimated by JO20. Its location in $(U-B)~vs~(V-B)$ TCD indicates that it does not meet the criteria of ZAMS within errors as depicted in Figure \ref{fig15}. Its distance is $2.014{\pm}0.348~kpc$ as per GAIA database, which is very far from the cluster~NGC1960. Thus, the variable star $V_{9}$ neither a main sequence star nor a member of cluster~NGC 1960. Its highest probabilistic values are just coincidence with the members of cluster. Since, its characteristics do not match well with those of a rotational type variable star, therefore author switched off its previous classification as did by JO20.\\
The slowly descending and quickly ascending nature of its phase-folded curve indicates that variable star $V_9$ is a RR Lyre star.\\
Star IDs 1431 ($5^{h}:36^{m}:45.65^{s}, ~+34^{0}:11^{'}:25.0^{''}$) and 1484 ($5^{h}:36^{m}:00.93^{s}, ~+34^{0}:09^{'}:07.7^{''}$) have been selected comparison stars for the variable star $V_{9}$. Their locations in Figures \ref{fig13} and \ref{fig12} are marked by C1(V9) and C2(V9) respectively. By visual inspection, author found that these comparison stars are far from the variable star $V_{9}$. The comparative light curves of star IDs 1431 and ID 1484 show a constant magnitude difference during entire observational session. According to GAIA database, the parallax values for Star Ids 1431 and 1484 are $0.6473~mas$ and $0.2435~ mas$, respectively. The colour-difference $(B-V)_0$ for both comparison stars is greater than 1.0 mag. Such pair of comparison stars is not possible for any given variable star as per differential photomery. Consequently, magnitude variation of stars is free from stellar distance and colour-difference in absolute/ standard photometry. Although, the author found some amount of magnitude variation for both comparison stars and this seems instrumental effect in nature.   
\subsection{Eclipsing Binaries type Variable Stars}
Eclipsing binaries are conveniently classified into two main groups: detached system or semidetached \citep{ba15}. Spherical or slightly ellipsoidal components are found in detached system (Algol type, EA), whereas tidally distorted stars are present in semidetached system ($\beta$ Lyre System, EB). The light remains nearly constant between eclipses of EA systems. Between eclipses, a continuous change of the combined brightness is found for EB system, making it impossibility to assign the exact times of onset and end of eclipses. Stars with planets can also show flux variations if associated planets for any star pass between Earth and the star.       
\subsubsection{Star IDs 900 ($V_5$), 902 and 904 of NGC~1960}
Star ID 902 is marked by $V_{55}$ by JO20 and assigned as a field star in their analysis. Furthermore, they found a minima in its phase curve with an amplitude  of $218~mmag$. In the case of $V_5(ID900)$ at figure 5, the different light curve show a partial eclipse and part of an eclipse. The full eclipse is detected in its light curve of 20 December 2013 and a portion of eclipse also detected in its light curve of 12 January 2015. The gap of both detected eclipses is  01 year 23 days 3 hours. The other nights show a constant data string. This is a classical light curve of an eclipsing binary due the rounded shape. Author found only one full minima in its light curves with amplitude of $248~mmag$ and it is close to estimated amplitude by JO20. In the present work, the light curves of variable star $V_{5}$ are shown in Figure \ref{fig05}. The time of eclipse is computed through the codes of 'PERIOD04' and 'PerSea' to be $0.3182{\pm}0.0848~d$ and $0.2857{\pm}0.0454~d$ respectively, which are not true periods. Both of these values are equivalent to the data strings, in which eclipse has detected. After visual inspection of light curves of this variable, estimated time of eclipse is approximately $0.216{\pm}0.018~d$. Its distance is $3.461{\pm}0.305~kpc$ and lies far away from the cluster periphery. Thus, it is a background for having field star in the field of cluster~NGC 1960.\\  
Light curves of Star IDs 902 ($5^{h}:36^{m}:25.02^{s}, ~34^{o}:09^{'}:17.5^{''}$) and 904 ($5^{h}:35^{m}:58.41^{s}, ~34^{o}:08^{'}:11.8^{''}$) have been selected for comparative analysis with those of the variable star $V_{9}$. Both comparison stars have nearby colour, ($(B-V)_0$), value with variable star $V_{9}$. The light curves of Star ID 904 has short periodic variation with low amplitude, whereas light curves of star ID 902 has irregular variation of brightness. Since, brightness fluctuation in their comparative light curves is almost constant, therefore, fluctuation of their light curves is a result of instrumental errors in nature.   
\subsubsection{Star IDs 1576, 1601 ($V_{11}$), 1613 of NGC~1960}
Based on the 'PERIOD04' and 'PerSea' code analysis, author obtained period values for $V_{11}$ as $1.1053{\pm}0.0007~d$ and $1.0980{\pm}0.1033~d$, respectively. Its distance from us is $1.230{\pm}0.053~kpc$ and its kinematic probability of membership in cluster is 0.73 as per proper motion via GAIA database. Thus, variable star $V_{11}$ is a member of cluster. It can be easily seen in its light curves, which are depicted in Figure \ref{fig07} that have short periodic variations superimposed with its principal period. The continuous change in brightness makes it difficult to assign the exact time of onset and end of eclipses of companion stars, and its phase-folded diagram shows the superimposed character of both eclipses as depicted in Figure \ref{fig11}. Hence, this star is classified as a semidetached eclipsing binary (EB) type variable star.\\
The pattern of brightness variation in light curves for star ID 1576 ($5^{h}:36^{m}:23.38^{s}, ~34^{o}:05^{'}:18.1^{''}$) is similar to that of variable star $V_{11}$ as depicted in Figure \ref{fig07}. As a result, author concluded that star ID 1576 is also an EB star with low amplitude. During the observational session of each night, light curves of star ID 1613 ($5^{h}:36^{m}:02.64^{s}, ~34^{o}:11^{'}:46.0^{''}$) shows an incremental slop over time with least inclination. In this connection, the fluctuation of its light curves is seem to be instrumental errors in nature.      
\subsection{Ellipsoidal Variable stars}
In the periodogram of an ellipsoidal binary, sharp peaks are found at the fundamental frequency and its harmonics. Usually only the first harmonic is visible and the amplitude of fundamental frequency is less than first harmonic in some cases. As a consequence of differences in harmonic content, the shapes of the light curves can be very different \citep{ba15}. These stars are close binary system and tidally distorted components. There are no eclipses in these variables due to the low inclination of the orbital axis, but the changing aspect towards us causes a change in brightness. Such brightness variation are a combination of tidal distortion, reflection and beaming. The period of the reflection and beaming contributions have the same period as the orbital period whereas the ellipsoidal effect has half the orbital period \citep{ba15}.    
\subsubsection{Star IDs 1123 ($V_6$), 1124 and 1130 of NGC~1960}
Based on the codes of 'PERIOD04' and 'PerSea', the period of $V_6$ is found to be $0.1528{\pm}0.0194~d$ and $0.1538{\pm}0.0369~d$, respectively. Its distance ($1.171{\pm}0.035~ kpc$) and kinematic probabilistic value ($0.765$) indicate that it is a member of cluster. Since, non-sinusoidal variation is found in its phase folded light curve, therefore it has the characteristics of ellipsoidal type variable star. However, a comprehensive analysis of its light curves during whole observational session indicates that it is an irregular type variable. As a result, it is classified as an irregular variable with character of ellipsoidal.\\
Star ID 1124 ($5^{h}:36^{m}:28.92^{s}, ~34^{o}:13^{'}:23.2^{''}$) has been selected as first comparison star for the variable star $V_6$. It shows almost constant magnitude with time as shown in Figure \ref{fig05}. Thus, it is classified as stable star in the direction of cluster NGC~1960 and its physical location is marked by C1(V6) in Figure \ref{fig12}.\\ 
Similarly, star ID 1130 ($5^{h}:35^{m}:48.78^{s}, ~34^{o}:07^{'}:46.3^{''}$) has been selected as second comparison star for the variable star $V_6$. This star is classified as an irregular type variable by JO20 and denoted by $V_{70}$ in their analysis. After visual inspection of its light curves in Figure \ref{fig05}, the author also confirms its character of irregular type variable star. 
\subsubsection{Star ID 1197, 1199 ($V_7$), 1204 of NGC~1960}
The value of period for variable star $V_7$ is computed to be $0.1747{\pm}0.0254~d$ and $0.1695{\pm} 0.0154~d$ using the codes of 'PERIOD04' and 'PerSea' respectively. In the present work, its kinematic probabilistic value is estimated to be 0.853 as per proper motion values via GAIA database. Unfortunately, its distance ($5.038 {\pm}0.698~kpc$) does not confirm its membership in the cluster, NGC~1960. Due to a non-sinusoidal variation of brightness in its light curves, variable star $V_{7}$ is classified as an ellipsoidal type variable star. Its position in $(U-B)$ vs $(B-V)$ TCD confirms its red character of colour.\\
Its First Comparison star, ID 1197 ($5^{h}:36^{m}:25.21^{s}, ~34^{o}:13^{'}:04.1^{''}$), shows nearly constant magnitude during observational session as depicted in Figure \ref{fig06}. As a result, it is a stable star and its position is marked by C1(V7) in Figure \ref{fig12}.\\ 
Similarly, light curves of its second comparison star, ID 1204 ($5^{h}:36^{m}:43.20^{s}, ~34^{o}:02^{'}:51.6^{''}$), show the character of irregular type variable star. 
\subsection{Rotational Variable}
Stellar rotation and magnetic activity are normally associated with a main-sequence star of G or later spectral type \citep{jo20}. The amplitude of pulsation of these stars is usually less than 0.1 mag. Thus, these stars are characterized by small amplitude, and red in colour $(B - V)_0 ~>~ 0.5 ~mag$. The periods of rotational variable stars can vary widely due to its tied with the own rotation of the stars.
\subsubsection{Star IDs 1345, 1348 and 1350 ($V_8$) of NGC~1960}
Using the codes of 'PERIOD04' and 'PerSea', the period values for $V_8$ are estimated to be $0.2864{\pm}0.0007~d$ and $0.2857{\pm}0.0947~d$ respectively. The lengths of individual data strings are $0.3167~d$ and $0.3000~d$ on date 20/12/2013 and 12/01/2015, respectively and these values are greater than $0.2864{\pm}0.0007~d$. Its character of variability is confirmed in individual data strings. It is located at $1.219{\pm}0.052~kpc$ from us and this value is close to the distance of cluster. In this connection, its kinematic probabilistic value for cluster membership is 0.761 leading member of studied cluster. It is red in colour, $(B-V)_{0}= 0.766~ mag$. Its location in $(U-B)$ vs $(B-V)$ TCD is found with ZAMS as shown in Figure \ref{fig15}. All the above characters make it a rotational type variable star.\\
The light curves of star IDs 1348 ($5^{h}:36^{m}:29.70^{s}, ~34^{o}:04^{'}:53.6^{''}$) and 1345 ($5^{h}:36^{m}:43.22^{s}, ~34^{o}:06^{'}:38.6^{''}$) are shown in Figure \ref{fig06}. Almost constant magnitude in light curves of star ID 138 make it a stable star, whereas star ID 1345 has a character of long periodic type variable stars.    
\subsubsection{Star IDs 1644 ($V_{12}$), 1732 and 1734 of NGC~ 1960}
Based on 'PERIOD04' and 'PerSea' code analysis, obtained values of period for variable star $V_{12}$ are $0.8538{\pm}0.0004~d$ and $0.6667{\pm}1.0505$ respectively. In $(U-B)$ vs $(B-V)$ TCD, it is located just on the main sequence. It has a red character in colour, $(B-V)_{0}= 0.946~mag$. As per GAIA database, its distance is found to be $1.103{\pm}0.049~kpc$ with a kinematic probability 0.791. Thus, it is a member of cluster, NGC~1960, along with the main sequence. The amplitude of this variable is found to be $94 ~mmag$ via code of PERIOD04. In this connection, it is classified as a rotational type variable star.\\
The light curves of star ID 1732 ($5^{h}:36^{m}:44.29^{s}, ~34^{o}:08^{'}:55.9^{''}$) shows short term variability with low amplitude, which appears to be instrumental error in nature. Beside this, it is approximately constant magnitude during observation. Other hand, the light curves of star ID 1734 ($5^{h}:36^{m}:46.46^{s}, ~34^{o}:12^{'}:51.3^{''}$) show a incremental slop of low inclination over observation as shown in Figure \ref{fig07}. 
\subsection{Low Amplitude Delta Scuti Variable}
Low-amplitude  delta Scuti stars (LADS) have pulsation with smaller amplitudes. The low-amplitude stars can be pre-main, main or post-main sequence stars, and can be either multiperiodic or monoperiodic \footnote{$aavso.org/vsots_delsct$}.
\subsubsection{Star IDs 1549 ($V_{10}$), 1553 and 1569 of NGC~1960}
The value of distance ($1.246{\pm}0.048~kpc$) and kinematic probability ($0.789$) of variable star $V_{10}$ confirm membership of the cluster. Its location in $(U-B)$ vs $(B-V)$ TCD makes it as a post-main sequence star. Author analyzed the frequencies of variable star $V_{10}$ with Fourier and variance analysis using codes of 'PERIOD04' and 'PerSea'. After these analyses,  the computed period is found to be $P_{1}=0.1886{\pm}0.0003$ and $P_{2}=0.2404{\pm}0.0226$, respectively, leading to $P_1/P_2=0.78$. The amplitude of pulsation is $74~mmmag$ as per Lomb-Scargle algorithm. So it is suggested that $V_{10}$ is a multi-periodic $\delta-Scuti$ star with low amplitude.\\
Due to almost constant difference of stellar magnitudes in the comparative light curves of Star IDs 1553 ($5^{h}:36^{m}:34.70^{s}, ~34^{o}:10^{'}:22.7^{''}$) and 1569 ($5^{h}:35^{m}:49.15^{s}, ~34^{o}:06^{'}:15.1^{''}$), their magnitude variation for one observational night is a result of instrumental transformation. However, the light curves of star ID 1569 show a character of long-term variability as shown in Figure \ref{fig07}.    
\subsubsection{Star IDs 2195, 2198 ($V_{15}$) and 2222 of NGC~1960}
The variable star ($V_{15}$) is a member of cluster due to its distance ($1.106{\pm}0.061 ~kpc$) and kinematic probability ($0.728$). It is also found to be a post-main sequence star according to its position in $(U-B)$ vs $(B-V)$ TCD. The values of period for variable star $V_{15}$ are computed using the codes of 'PERIOD04' and 'PerSea' to be $0.3039{\pm}0.0810~d$ and $0.2041{\pm}0.0105~d$ using the codes of 'PERIOD04' and 'PerSea' respectively. Since, the computed value of $0.3039{\pm}0.0810~d$ (as per PERIOD04) is close to the length of individual data string therefore, it is not a true period for variable star $V_{15}$. However, it leads to a ratio of $\nu_{1}:\nu_{2}=3:2$. As per Lomb-Scargle algorithm, the pulsation amplitude for this star is $72~mmag$. Thus, it appears to be low amplitude $\delta-Scuti$ type variable star.\\
The light curves of star IDs 2195 ($5^{h}:36^{m}:08.50^{s}, ~34^{o}:07^{'}:37.9^{''}$) and 2222 ($5^{h}:35^{m}:55.14^{s}, ~34^{o}:10^{'}:10.8^{''}$) show no sign of reasonable variability and their comparative light curves show almost constant difference for their magnitudes as shown in Figure \ref{fig08}.  
\subsection{Irregular and miscellaneous Variable Stars}
Mira variables have less regular light curves with large amplitudes of several orders of magnitudes, while semi-regular variables have less regular with smaller amplitudes \citep{sa17}. In this connection, the amplitudes of light curves of irregular variable stars do not occur after a fixed time interval and shape of their light curves has been found to be in an uncertain pattern. Consequently, the variations in brightness show no regular periodicity in an irregular type variable star. Such stars are divided into eruptive and  irregular pulsating variable stars. The variation of brightness of an eruptive variable star happens due to violent processes and flares occurring in its chromosphere ana coronae. Eruptive variable stars are found near the main sequence.   
\subsubsection{Star IDs 635, 645 and 649 ($V_3$) of NGC~1960}
The codes of 'PERIOD04' and 'PerSea' give its period value as $0.3598{\pm}0.0001~d$ and $0.4000{\pm}0.2086~d$ respectively. These values are close to length of individual data strings in 2015. The length of data string ($0.3598{\pm}0.0001~d$) of variable star $V_3$ lead to a peculiar phase-fold diagram with slowly descending and quick ascending branches as similar to the $RR~Lyre$ variable stars. In this connection, its least kinematic probabilistic value and distance indicate that it is not a member of cluster NGC\,1960. However, its position in $(B-V) ~vs~V$ CMD and $(U-B)~vs~(V-B)$ TCD is just on the main sequence.  Its position in $(U-B)~vs~(V-B)$ TCD is found to be far away from the group of identified $\delta-Scuti$ variable stars. In addition, length of individual data strings is too long for normal $\delta~Scuti$ star. Under these circumstance, it is impossible to classify its variability type. As a result, it is listed as miscellaneous variable star in present analysis.\\
In the case of this variable star, Star IDs 635 ($5^{h}:36^{m}:09.47^{s}, ~+34^{0}:08^{'}:51.2^{''}$) and 645  ($5^{h}:36^{m}:29.11^{s}, ~+34^{0}:07^{'}:42.2^{''}$) are selected for the comparative analysis. Star ID 635 is assigned as a $\gamma-Dor$ variable star by JO20. In this connection, author found that light curves of Star ID 635 show characteristic similarities to that of star ID 624. Similarly, light curves of star IDs 616 and 645 have similar characteristics. Thus, the nature of the light curves of star IDs 616, 624, 635 and 645 are either uncertain in nature or have some kind of irregular variability.
\subsubsection{Star IDs 1701, 1702 ($V_{13}$) and 1732 of NGC~1960}
The distance for variable star, $V_{13}$, is $2.269{\pm}0.188~kpc$ via proper motion values as extracted from the GAIA database. Its kinematic probability coincides with that of the member of cluster. Its light curves show a non-sinusoidal variation of brightness as shown in Figure \ref{fig08}. As per codes of 'PERIOD04' and 'PerSea', the estimated values of period for variable star $V_{13}$ are $0.3057{\pm}0.0815~d$ and $0.2857{\pm}0.0762~d$, respectively. Both values are similar to time length of individual data string during 2015, which is a misleading period. Phase curve via $0.3057{\pm}0.0815~d$ poses a fine curve as shown in Figure 11 for ID 1702. This period coincidentally seems close to length of the data string. As a result, it has been classified as miscellaneous type variable star due to lack of continuous data.\\
Star IDs 1701 ($5^{h}:36^{m}:36.95^{s}, ~34^{o}:11^{'}:57.0^{''}$) and 1732 ($5^{h}:36^{m}:44.29^{s}, ~34^{o}:08^{'}:55.9^{''}$) are selected for comparative analysis and their positions are marked by C1(V13) and C2(V13) in Figure \ref{fig13}. After visual inspection of light curves of both comparison stars in Figure \ref{fig08}, author found a constant magnitude difference between the both comparison stars. As a result, it is concluded that character of irregular type variability of both stars is an instrumental effect in nature.
\subsubsection{Star IDs 1893, 1914 ($V_{14}$) and 1915 of NGC~1960}
Using codes of 'PERIOD04' and 'PerSea', estimated value of period for variable star $V_{14}$ is computed to be $0.2665{\pm}0.0711~d$ and $0.2222{\pm}0.0775~d$, respectively. Both computed values are close to average length of data strings. It is located at $10.050{\pm}14.555~kpc$ as per extracted parallax from GAIA database, which is highly uncertain. A high kinematic probability is not enough to assign membership status to a star. Its location in $(U-B)$ vs $(B-V)$ TCD is just on the main sequence. Its depicted light curves in Figure \ref{fig08} show a non-sinusoidal variation of brightness. The pixel coordinates and light curves of this variable star suggested that its flux is affected by instrumental errors leads an uncertainty for computing periods as well as analysis of its light curve. As a result, it is classified as miscellaneous type variable star. \\
In Figure \ref{fig13}, the star IDs 1893 ($5^{h}:36^{m}:13.78^{s}, ~34^{o}:10^{'}:18.2^{''}$) and 1915 ($5^{h}:36^{m}:18.07^{s}, ~34^{o}:13^{'}:58.4^{''}$)  are marked by C1(V14) and C2(V14) respectively. The light curves of Star ID 1893 show the character of a long periodic variables, whereas Light curves of star ID 1915 are almost constant for some night of observation and have incremental slope with low inclination for other nights of observation. 
\subsubsection{Star IDs 2323 ($V_{16}$), 2352 and 2379 of NGC~1960}
Using codes of 'PERIOD04' and 'PerSea', the estimated values of period for $V_{16}$ are $0.3005{\pm}0.0801~d$ and $0.2941{\pm}0.0171~d$ respectively. Its phase curve via $0.3005{\pm}0.0801~d$ posses a character of asymmetrical magnitude variation with increasing rapidly and decreasing slowly. The said phase curve seems to be distorted in nature as depicted in Figure \ref{fig11}. Both periods were found close to length of individual data-string in 2015, therefore both are not true period for variable star. Such misleading period produces distortion in phase diagram compared to their light curves. Its distance is $0.864{\pm} 0.036~kpc$ according to parallax value as extracted from GAIA database and it is close to us with respect to the cluster NGC~1960. In this connection, its least kinematic probability (as per GAIA database) also confirms it as a field Star. Although, its position in $(U-B)~vs~(V-B)$ TCD is found just on the main sequence as depicted in Figure \ref{fig15}.\\
Author has selected two stars, ID 2352 ($5^{h}:36^{m}:03.82^{s}, ~+34^{0}:11^{'}:51.5^{''}$) and ID 2379 ($5^{h}:36^{m}:12.86^{s}, ~+34^{0}:07^{'}:54.1^{''}$), to compare their light curves with those of variable star $V_{16}$. Both stars show approximate constant magnitude difference during entire observation sessions. Although, their light curves have character of Long Periodic Variables, but seems to be instrumental effect. As per observations of each individual night, their magnitude variation less than that of the variable star $V_{16}$ as shown in upper panels of Figure \ref{fig09}. The pixel distances of these stars (IDs 2323, 2352, 2379) indicate that they are far from each other, as listed in Table 6. Colour, $(B-V)_o$, values of $V_{16}$ and ID 2352 are close to each other, but away from that of ID 2379. Thus, the similar character of light curves for ID 2352 and ID 2379 confirms colour- free selection of comparison stars during time series analysis via absolute photometry.   
\subsubsection{Star IDs 2439, 2451 ($V_{17}$) and 2479 of NGC~1960}
The values of distance and kinematic probability for the variable star $V_{17}$ are found to be $1.107{\pm}0.055 ~kpc$ and $0.771$ respectively. Consequently, it is assigned as a member of cluster and found near the main sequence as shown in Figure \ref{fig15}. Author did not find any sign of regular pulsation for it. A speck of stellar brightness for this variable star has detected on date 11 December 2013 as depicted in Figure 9. The period of this variable star can not be computed by the codes of 'PERIOD04' and 'PerSea'. Thus, this variable star is suggested to be an irregular type variable.\\
In the case of variable star $V_{17}$, the comparison stars are ID~2439 ($5^{h}:36^{m}:14.60^{s}, ~34^{o}:07^{'}:21.1^{''}$) and ID 2479 ($5^{h}:35^{m}:56.35^{s}, ~34^{o}:05^{'}:53.3^{''}$). The comparative light curves of both comparison stars show almost constant difference of brightness during observation of each night. Comparative light curves of $V_{17}$ with its comparison stars indicate that spacing between their light curves varies night to night as depicted in the middle Set of Panels of Figure 9. It may be due to character of long variability of $V_{17}$ with irregular pulsation. In addition, the light curves of star ID 2439 also show a character of long term variability. 
\subsubsection{Star IDs 2868, 2875 ($V_{18}$) and 2889 of NGC~ 1960}
The variable star $V_{18}$ is a cluster member due to its kinematic probability $(0.764)$ of membership and distance $(1.191{\pm}0.065 ~kpc)$. The light curves of variable star $V_{18}$ show no sign of periodic pulsation. As per observation on date on date 11 December 2013, flare type structure has been detected in the constructed light curve of this variable star. Its location is found near the main sequence in $(U-B)$ vs $(B-V)$ TCD and depicted in Figure \ref{fig15}. The codes of 'PERIOD04' and 'PerSea' are not able to compute period for this variable star. Consequently, it is classified as an irregular variable star with eruptive phenomena.\\
In the case of $V_{18}$, the comparison stars are ID 2868 ($5^{h}:36^{m}:34.54^{s}, ~34^{o}:08^{'}:31.6^{''}$) and ID 2889 ($5^{h}:35^{m}:55.53^{s}, ~34^{o}:07^{'}:52.2^{''}$). Due to the almost constant brightness in the light curves for star ID 2868, it is assigned as a stable star. The black comparative light curves of field stars shows no noticeable fluctuation in flux, whereas, comparative light curves of $V_{18}$ variable and field stars confirm the long term stellar variability of $V_{18}$ as depicted in the lower set of panels of Figure 9.
\subsection{Characteristics test of $\delta-Scuti-\gamma-Doradus$ hybrid variables}
Generally, $\gamma-Doradus$ instability strip is found below the $\delta-Scuti$ strip. Several authors pointed out that some portion of instability strips of both these classes overlap each other. Such hybrid candidates have already been discovered in open clusters by \cite{ha08} and \cite{jo20}. The estimated log(age) of cluster NGC~1960 is $7.35{\pm}0.05~yr$ and found close to J20. It belongs to a young open cluster in which the A/F-type stars have not evolved. $\delta-Sct$ stars have high pulsation mode frequencies ($\nu \geq 35~d^{-1}$) as inferred from theoretical models \cite{mi17}. Such fast variability should be detectable in the light curves, except for the fact that the amplitudes are very small.

\subsubsection{Star IDs 592, 595 and 600 ($V_1$) of NGC~1960}
\begin{figure*}
\includegraphics[width=16.0cm]{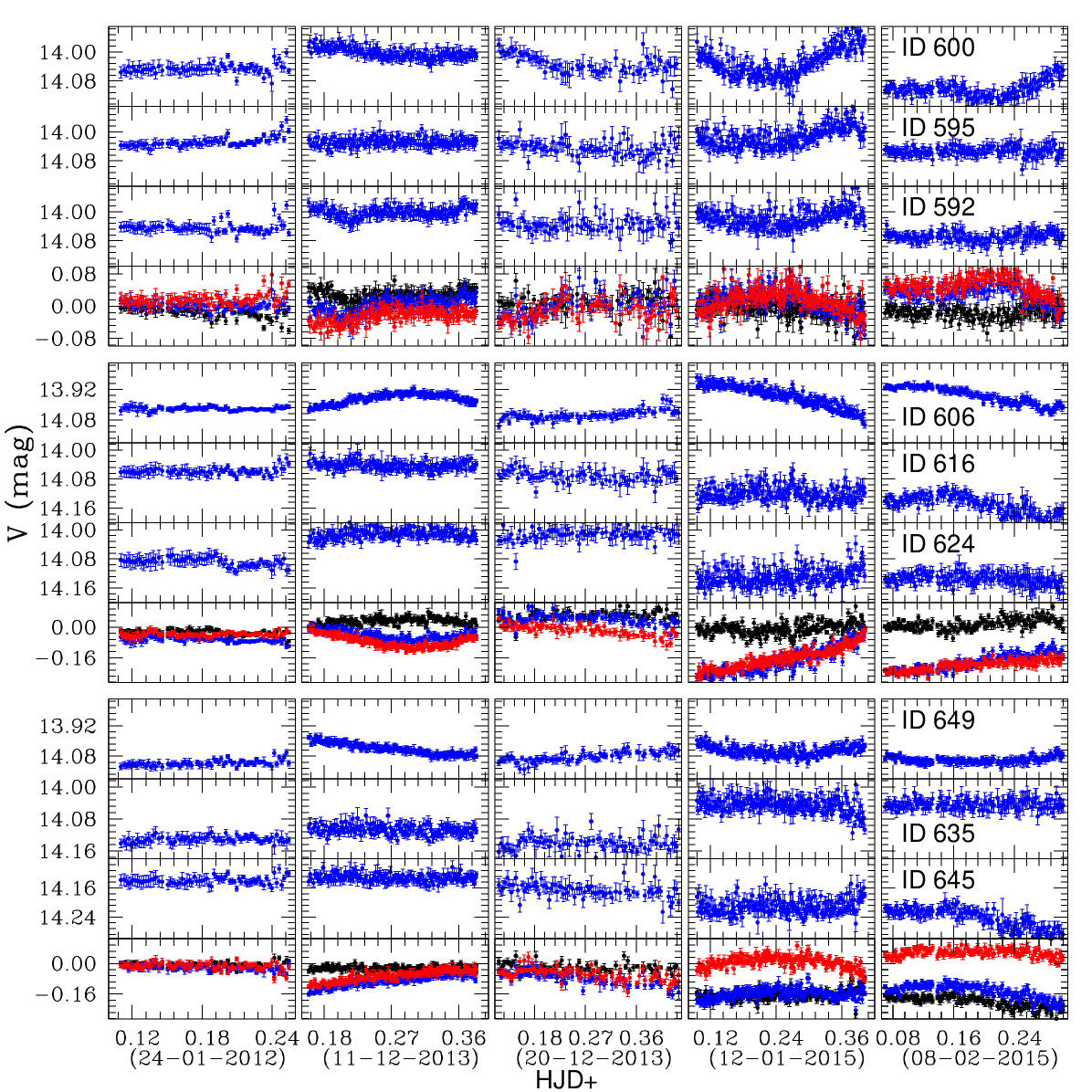}
\caption{The panels represent the stellar light curves for star IDs 592, 595, 600, 606, 616, 624, 635, 645 and 649  of cluster NGC 1960. The star IDs 592 and 595 are selected comparison stars for Potential variable ($V_{1}$, Star ID 600). The star IDs 616 and 624 are selected comparison stars for Potential variable ($V_{2}$, Star ID 606). The star IDs 635 and 645 are selected comparison stars for Potential variable ($V_{3}$, Star ID 649).}
\label{fig04}
\end{figure*}

\begin{figure*}
\includegraphics[width=16.0cm]{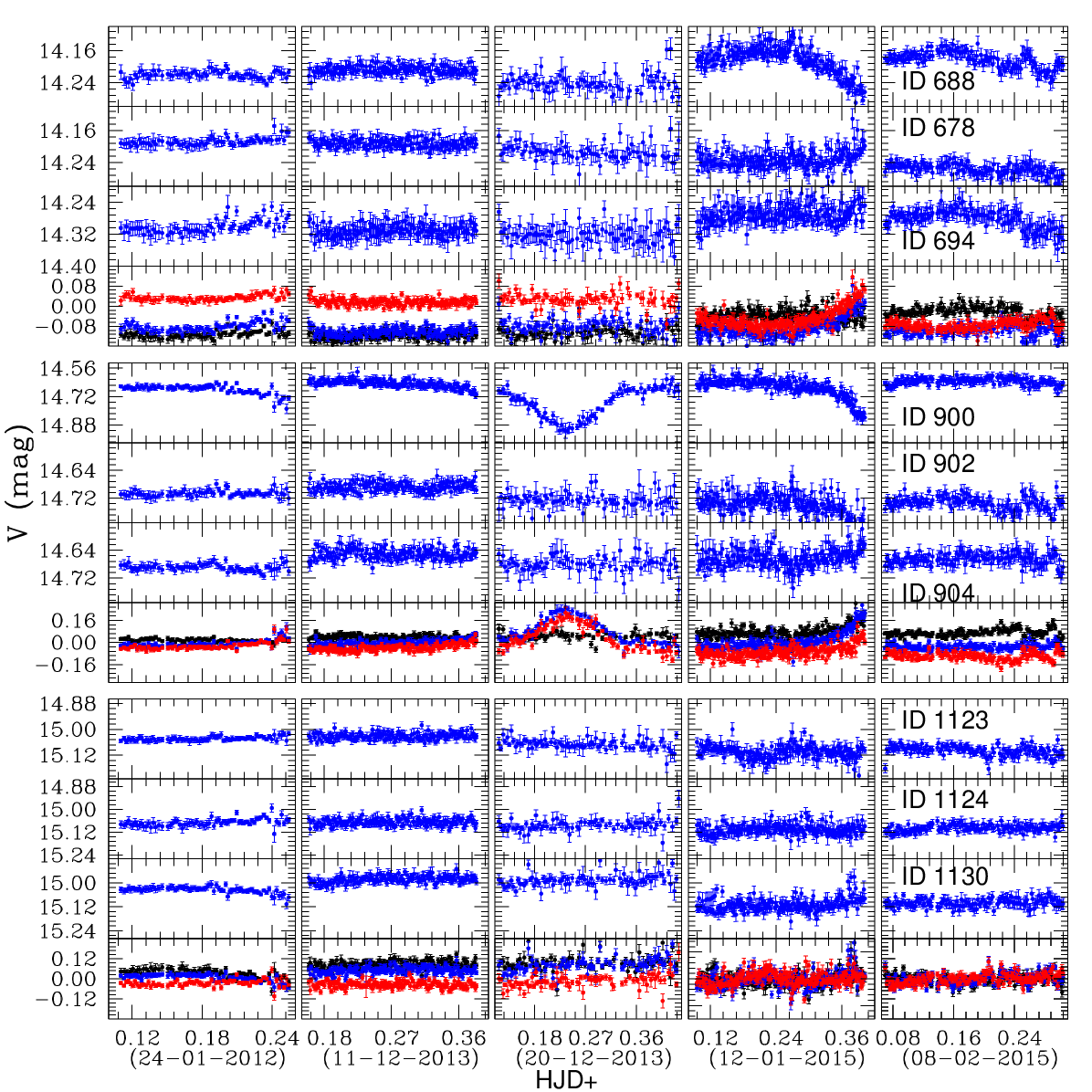}
\caption{The panels represent the stellar light curves for star IDs 678, 688, 694, 900, 902, 904, 1123, 1124 and 1130  of cluster NGC 1960. The star IDs 678 and 694 are selected comparison stars for Potential variable ($V_{4}$, Star ID 688). The star IDs 902 and 904 are selected comparison stars for Potential variable ($V_{5}$, Star ID 900). The star IDs 1124 and 1130 are selected comparison stars for Potential variable ($V_{6}$, Star ID 1123).}
\label{fig05}
\end{figure*}

\begin{figure*}
\includegraphics[width=16.0cm]{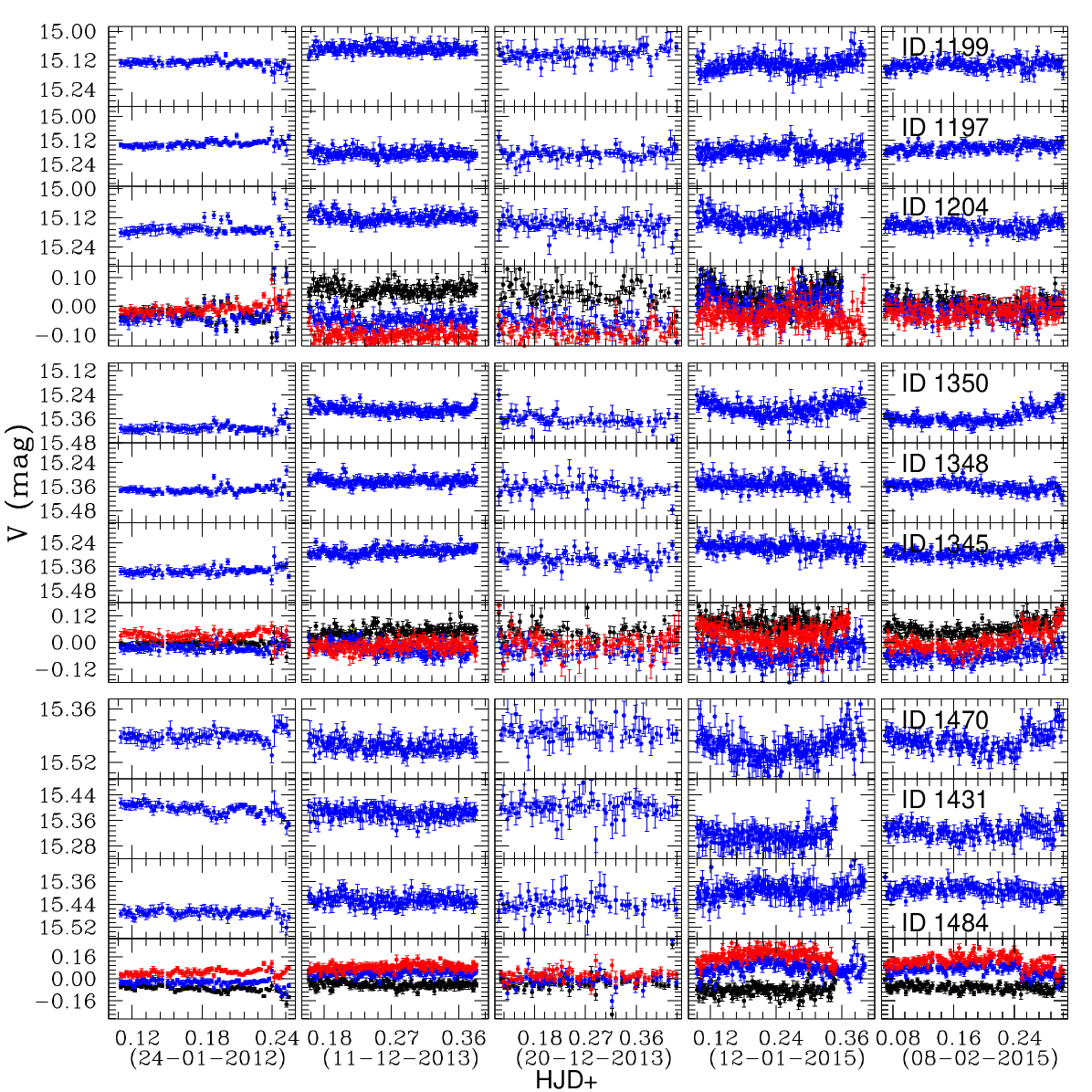}
\caption{The panels represent the stellar light curves for star IDs 1197, 1199, 1204, 1345, 1348, 1350, 1431, 1470 and 1484  of cluster NGC 1960. The star IDs 1197 and 1204 are selected comparison stars for Potential variable ($V_{7}$, Star ID 1199). The star IDs 1345 and 1348 are selected comparison stars for Potential variable ($V_{8}$, Star ID 1350). The star IDs 1431 and 1484 are selected comparison stars for Potential variable ($V_{9}$, Star ID 1470).}
\label{fig06}
\end{figure*}

\begin{figure*}
\includegraphics[width=16.0cm]{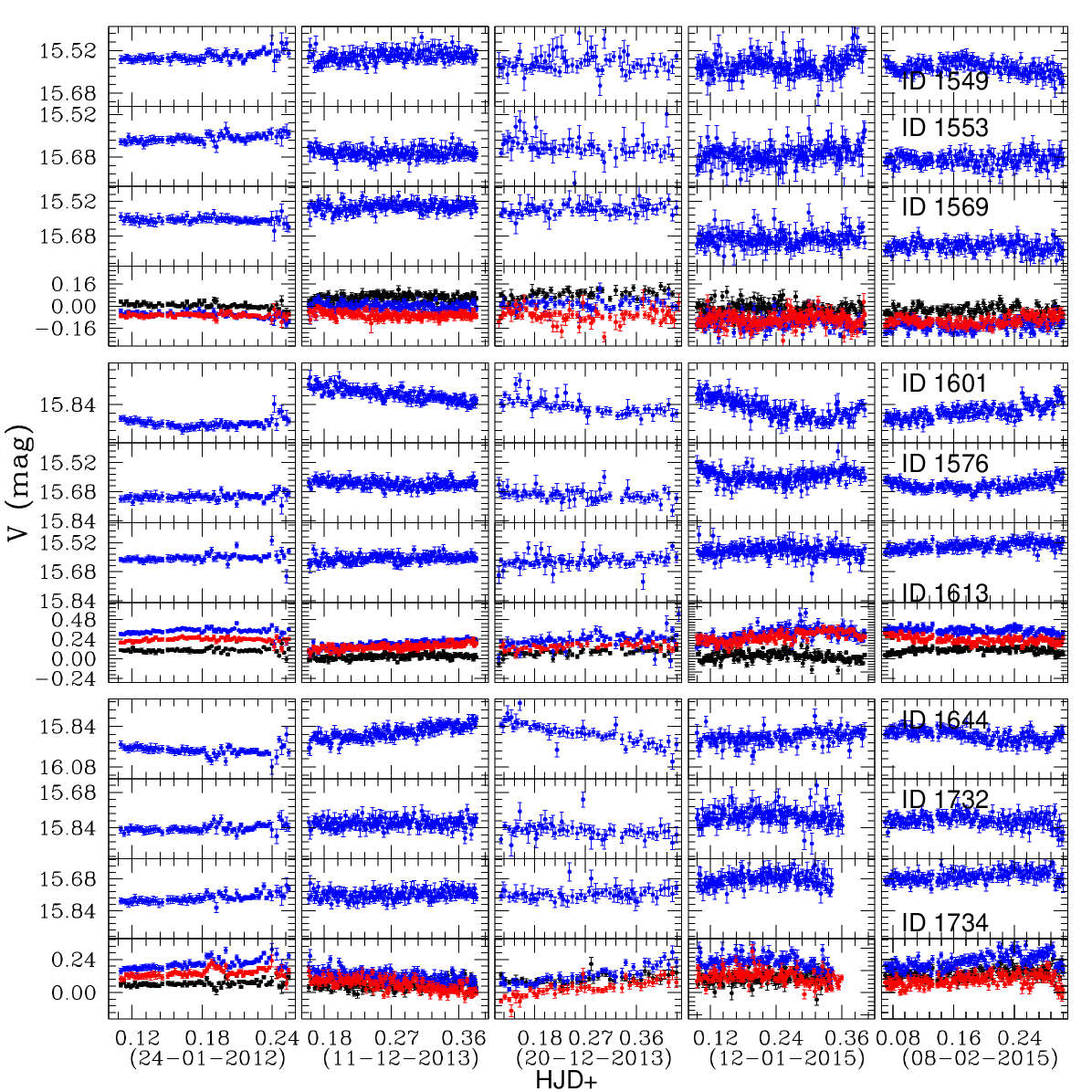}
\caption{The panels represent the stellar light curves for star IDs 1549, 1553, 1569, 1576, 1601, 1613, 1644, 1732 and 1734 of cluster NGC 1960. The star IDs 1553 and 1569 are selected comparison stars for Potential variable ($V_{10}$, Star ID 1549). The star IDs 1576 and 1613 are selected comparison stars for Potential variable ($V_{11}$, Star ID 1601). The star IDs 1732 and 1734 are selected comparison stars for Potential variable ($V_{12}$, Star ID 1644).}
\label{fig07}
\end{figure*}

\begin{figure*}
\includegraphics[width=16.0cm]{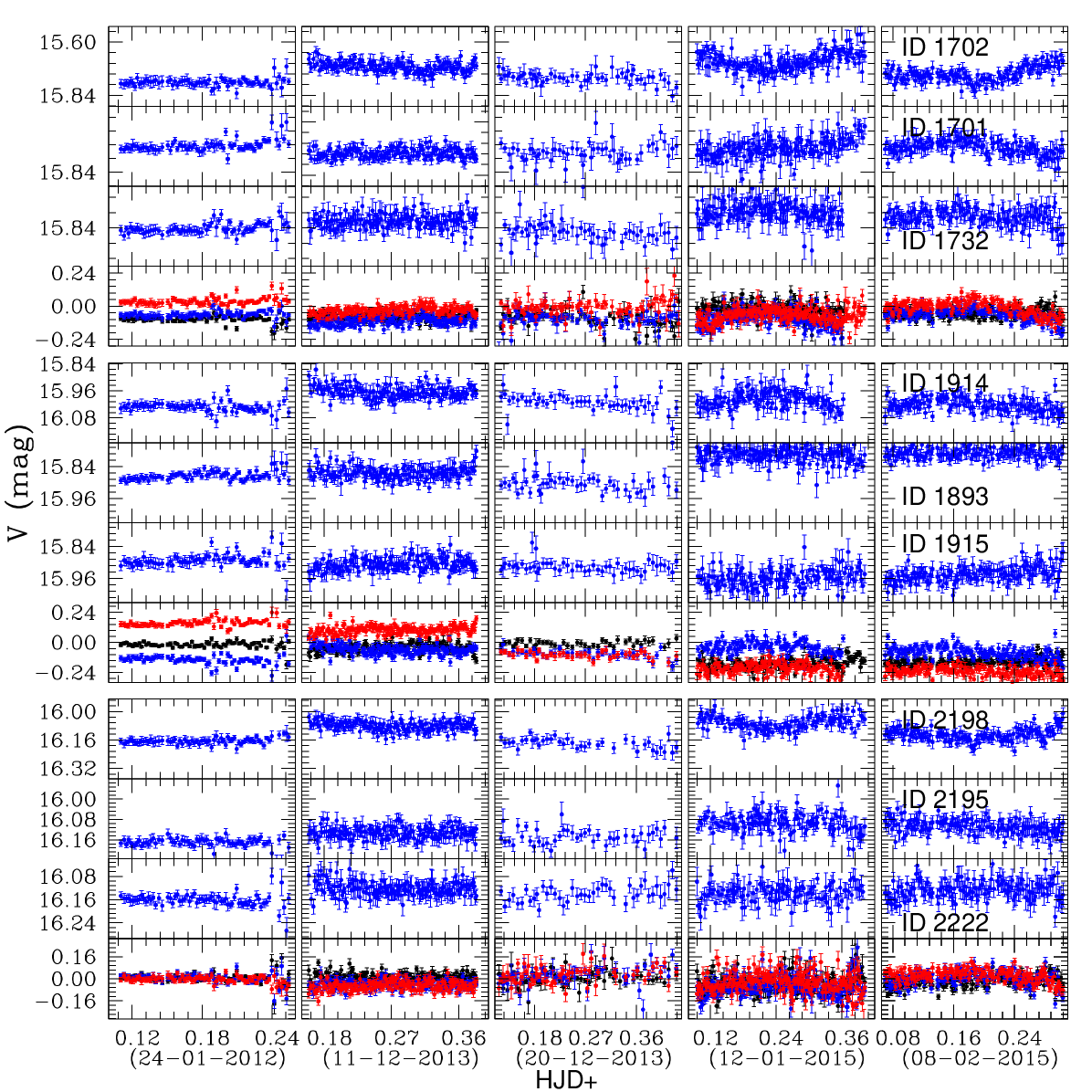}
\caption{The panels represent the stellar light curves for star IDs 1701, 1702, 1732, 1893, 1914, 1915, 2195, 2198 and 2222 of cluster NGC 1960. The star IDs 1701 and 1732 are selected comparison stars for Potential variable ($V_{13}$, Star ID 1702). The star IDs 1893 and 1915 are selected comparison stars for Potential variable ($V_{14}$, Star ID 1914). The star IDs 2195 and 2222 are selected comparison stars for Potential variable ($V_{15}$, Star ID 2198).}
\label{fig08}
\end{figure*}

\begin{figure*}
\includegraphics[width=16.0cm]{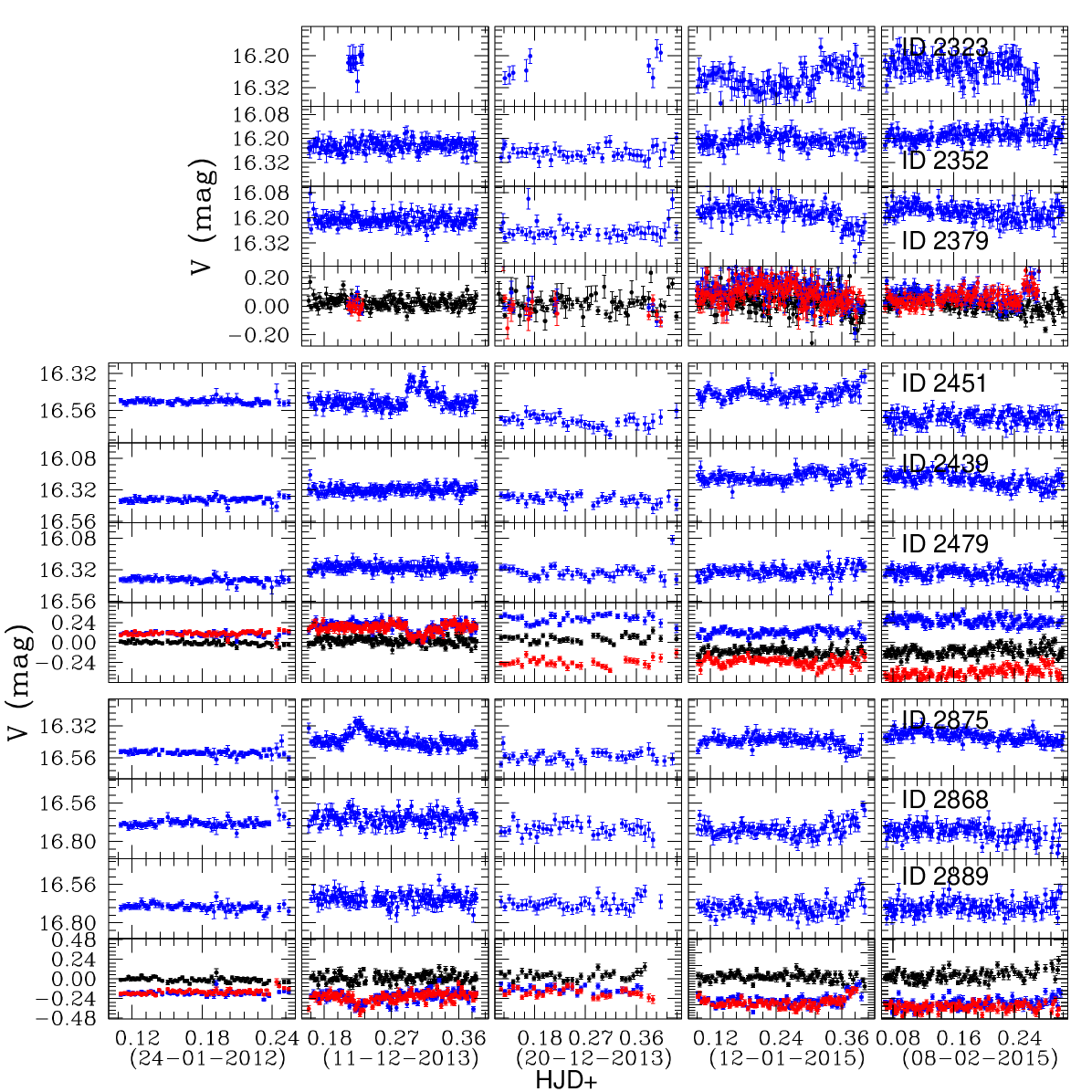}
\caption{The panels represent the stellar light curves for star IDs 2323, 2352, 2379, 2439, 2451, 2479, 2868, 2875 and 2889 of cluster NGC 1960. The star IDs 2352 and 2379 are selected comparison stars for Potential variable ($V_{16}$, Star ID 2323). The star IDs 2439 and 2479 are selected comparison stars for Potential variable ($V_{17}$, Star ID 2451). The star IDs 2868 and 2889 are selected comparison stars for Potential variable ($V_{18}$, Star ID 2875).}
\label{fig09}
\end{figure*}

\begin{figure*}
\centerline{\includegraphics[width=16.0cm]{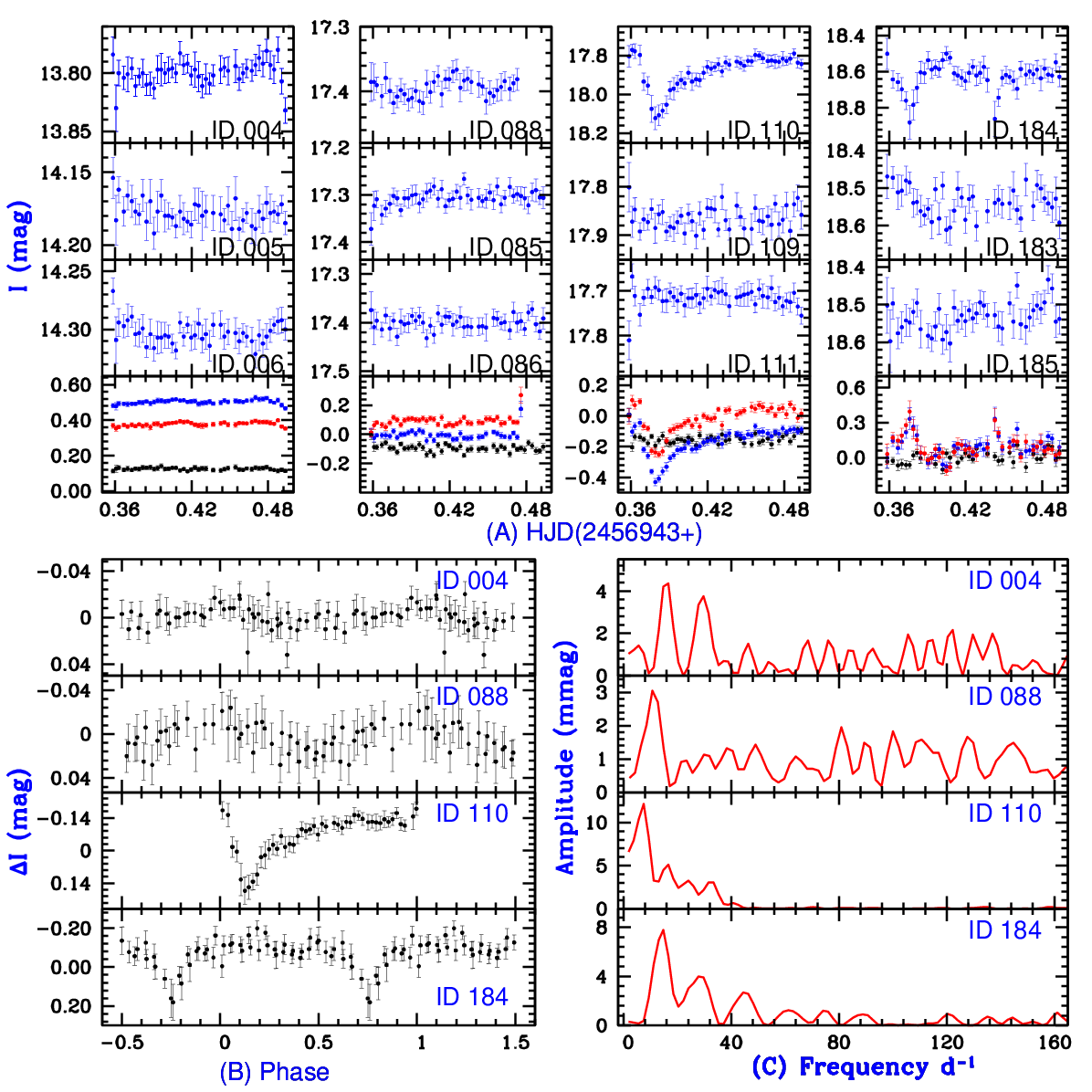}}
\caption{{\bf (A):-} We represent the light curves of identified varibles (ID 004, ID 088, ID 110, ID 184) and their correspanding comparision stars in the field-view of DOLIDZE 14. The HJD time of observations are shown in x-axis whereas y-axis is shown apparent magnitudes of stars in $I$-filter. {\bf (B):-} The panels show light-pholded-curves or phase diagrams of identified variables. The value of phase and amplitude ($mmag$) of stellar variability are shown in the x-axis and y-axis respectively. {\bf (C):-} The frequency spectrum of identifed varaibles of DOLIDZE 14 are depicted here. The frequency ($d^{-1}$) and amplitude ($mmag$) of variables are represented in x-axis and y-axis respectively.\label{fig10}}
\end{figure*}

\begin{figure*}
\includegraphics[width=16.0cm]{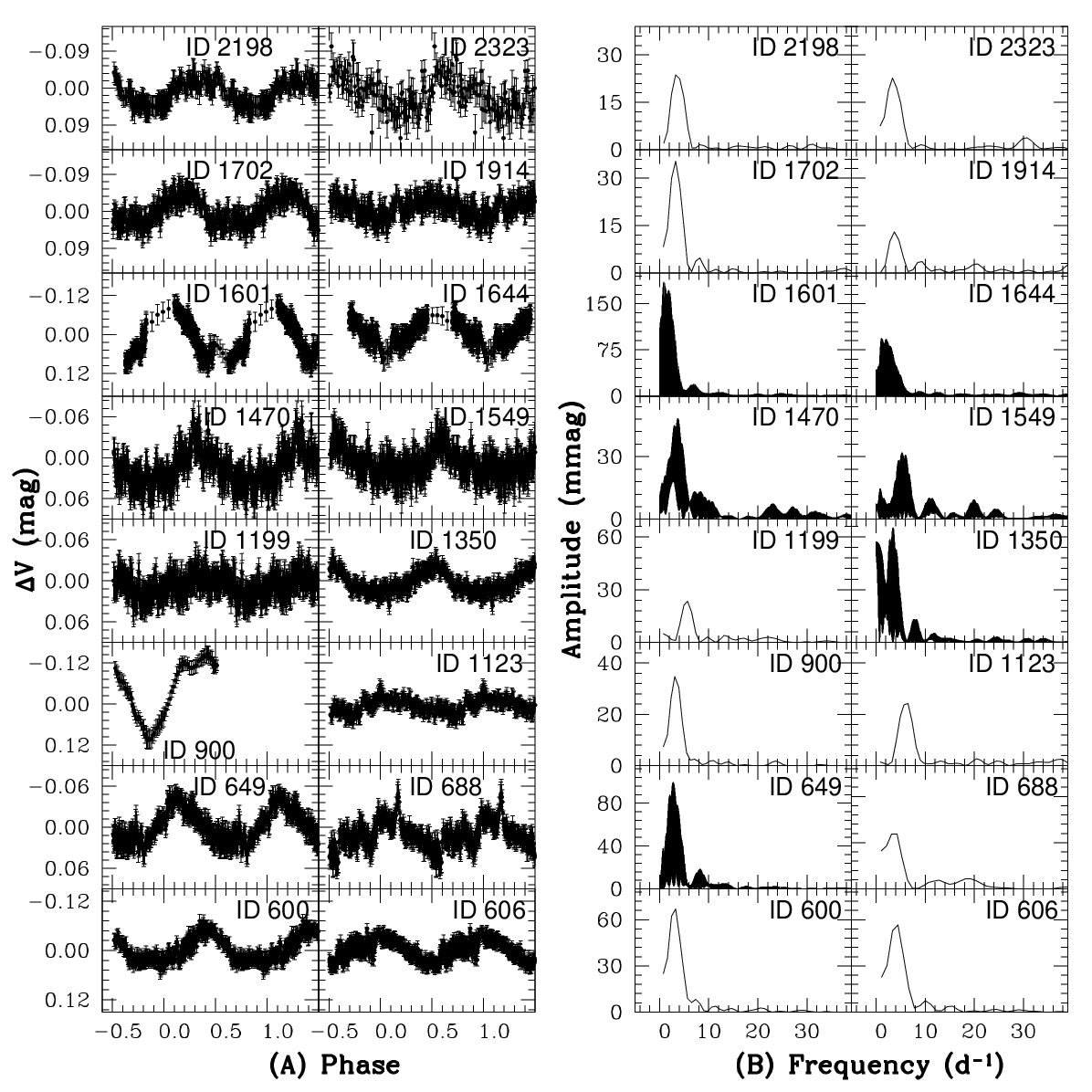}
\caption{The left panels represent the phase-folded-diagrams of identified variable stars within the cluster NGC 1960, whereas their corresponding DFT represent in the right panels.}
\label{fig11}
\end{figure*}

\begin{figure*}
\includegraphics[width=16.0cm]{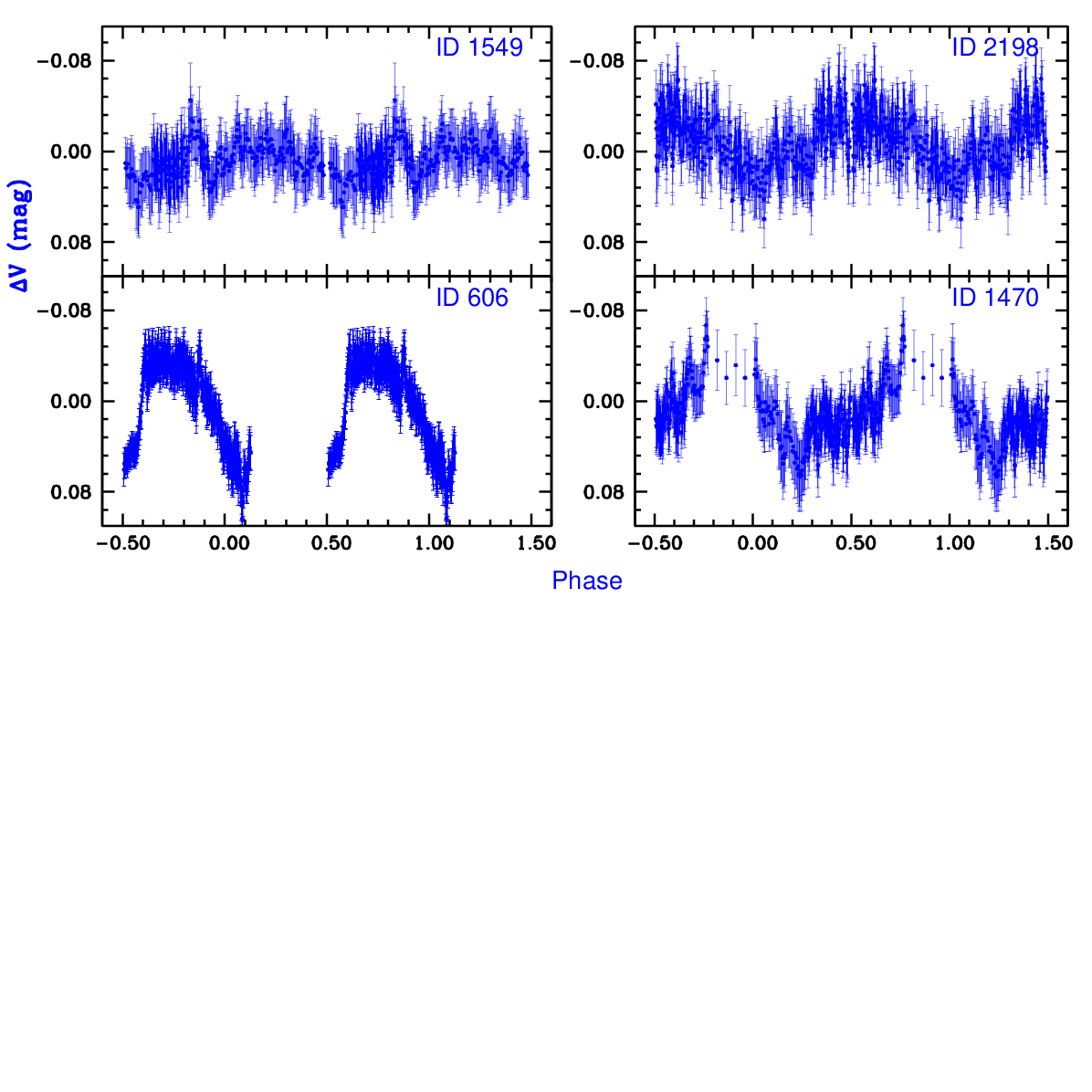}
\vspace{-7cm}
\caption{The panels represent the phase-folded-diagrams of variable stars of cluster NGC 1960 as per period via ANOVA.}
\label{fig12}
\end{figure*}

Author has selected two stars, ID 592 ($5^{h}:36^{m}:39.44^{s}, ~+34^{0}:06^{'}:12.3^{''}$) and ID 595 ($5^{h}:36^{m}:41.27^{s}, ~+34^{0}:09^{'}:29.9^{''}$), to compare their light curves with the Star ID 600 and these stars are marked by C1(V1) and C2(V2) in Figure~\ref{fig13}. \cite{jo20} assigned ID 595 as a cluster member and a hybrid $\delta~Scuti~\gamma~Daradus$ variable star and denoted it by $V_{27}$ in their analysis. It has the lowest V-magnitude and colour (B-V) of all stars, listed in Table 5. Upon inspection of light curves of Star IDs 592 and 595 in Figure~\ref{fig04}, author noted that both stars showed a approximate similar pattern. The light curves of ID 595 have a regular pattern of flux variation with very low amplitude and author considered it as an effect of instrumental pseudo-variability. Thus, it is impossible to find character of young $\delta-Scuti$ stars based on low quality data sets. The light curves of star ID 600 have shown prominent variation in its magnitude. Its period is found to be $0.3057{\pm}0.0815~d$ and $0.4000{\pm}0.1055~d$ by the codes of 'PERIOD04' and 'PerSea', respectively. These value are similar to data strings. As a result, author did not advocated about its true period. It has characteristics similar to the variable star $V_4$ star and it is listed as miscellaneous in the present analysis. Its location within cluster is shown by the mark M36(V1) in Figure~\ref{Fig2}. It is also found to be more kinematic probabilistic member of the cluster.
\subsubsection{Star IDs 678, 688 ($V_4$) and 694 of NGC~1960)}
\begin{figure*}
\includegraphics[width=16.0cm]{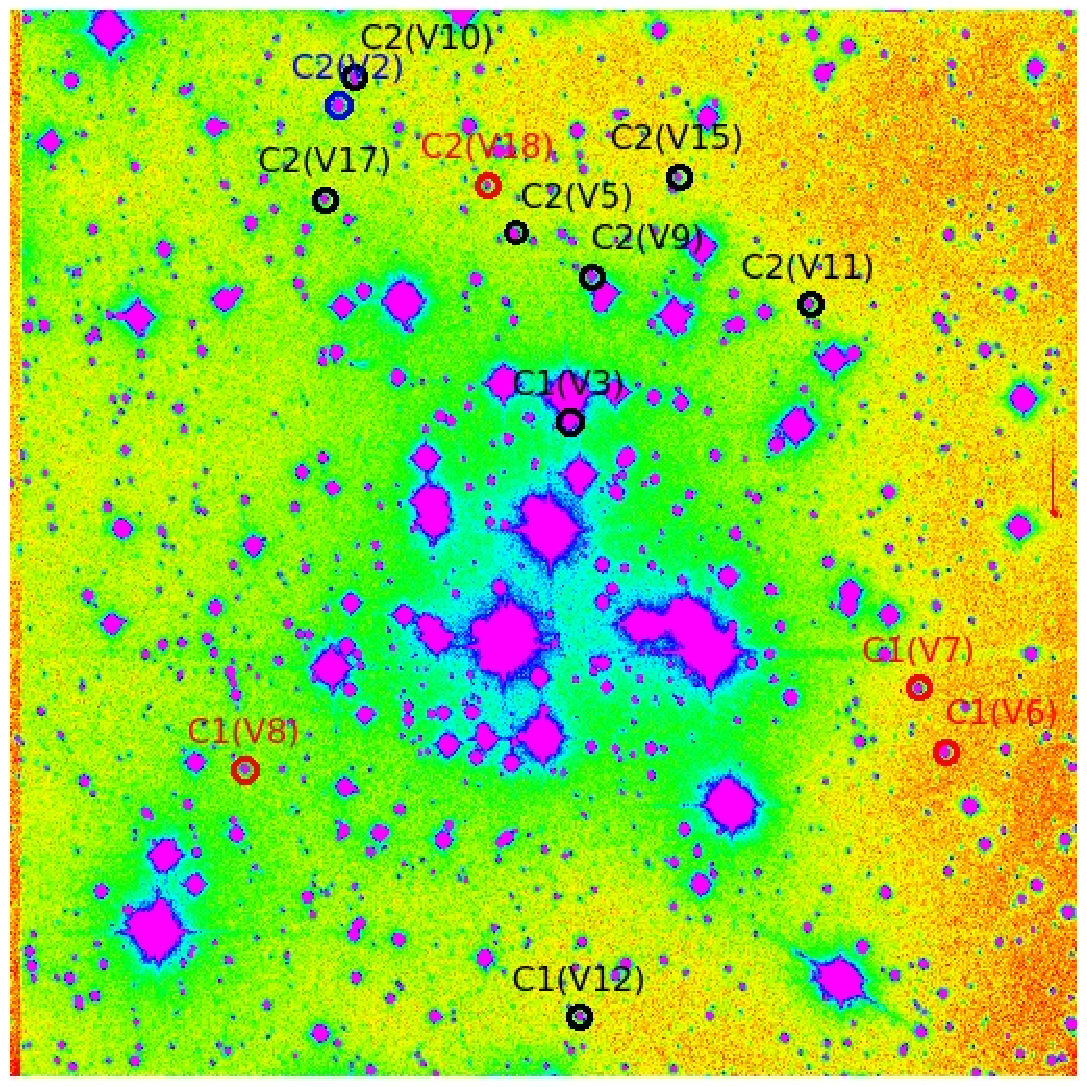}
\caption{The finding chart for selected comparison stars within cluster NGC~1960. Identified standard stars are marked by red open circles and they have almost constant brightness/flux during observations. Black open circles represent those long periodic variable stars, for which approximate flux is found for a particular night of observation. However, the value of magnitude is varying nigh to night.}
\label{fig13}
\end{figure*}

Star IDs 678 ($5^{h}:36^{m}:33.38^{s}, ~+34^{0}:10^{'}:13.0^{''}$) and 694 ($5^{h}:36^{m}:37.21^{s}, ~+34^{0}:12^{'}:39.3^{''}$) are selected as possible comparison star. These stars are marked by C1(V1) and C2(V2) in Figure~\ref{fig13} and their physical coordinates are found closest compare than other set of stars. Variable star $V_{4}$ is assigned to be a $\gamma~Daradus$ variable star with period $1.07066~d$ and marked by $V_{62}$ in the work of \cite{jo20}. In the present analysis, its period is found to be $0.3115{\pm}0.1168~d$ by the code of 'PERIOD04', whereas the said value is $0.2857{\pm}0.0706$ by the ANOVA analysis as per code of 'PerSea'. These values are closed to average length of data strings. It is more kinematic probabilistic member of cluster and marked by M36(V4) in Figure~\ref{Fig2}. In present study, author did not classify it due to circumstances of data. 

 The trend of light curves of Star IDs 678 and 694 are approximately similar to each other. However, light curves of star ID 694 shows a regular feature of short periodic variability and it is considered to be pseudo-variability by author.
\section{Results and Discussion}
The contamination of fluxes of fainter stars is occurred due to their nearby brighter stars. Author also found very high scattering of data points in the light curves of brighter stars of NGC\,1960 due to their saturation in the present deep photometric observations. As a result, author did not analysis the time series observations of bright and their nearby stars.

Star IDs 1124, 1197, 1348 and 2868 of M36 have approximately constant stellar magnitude during whole observations, their celestial coordinates are ($5^{h}:36^{m}:28.92^{s}, ~34^{o}:13^{'}:23.2^{''}$), ($5^{h}:36^{m}:25.21^{s}, ~34^{o}:13^{'}:04.1^{''}$), ($5^{h}:36^{m}:29.70^{s}, ~34^{o}:04^{'}:53.6^{''}$) and ($5^{h}:36^{m}:34.54^{s}, ~34^{o}:08^{'}:31.6^{''}$) respectively. Thus, these stars are classified as the constant stars within cluster. According to the GAIA database, the distance of stars 1124, 1197, 1348 and 2868 has been computed as $1.166{\pm}0.035~kpc$, $1.152{\pm}0.036~kpc$, $3.461{\pm}0.278~kpc$ and $1.419{\pm}0.109~kpc$ respectively. Thus, star IDs 1124 and 1197 are members of cluster NGC~1960.

In the present analysis, some variable stars (such as $V_2$) show two distinct period values via two distinct algorithm (ANOVA and statistical analysis. In the case of variable star $V_2$, period value via statistical analysis is close to the length of individual data strings in 2015, therefore it is not suitable to consider a true period. As a result, true period for variable star $V_2$ is $0.6250{\pm}0.0274~d$ as per ANOVA analysis.

Among the 18 detected variable stars, only four are cross matched with JO20, whereas three of the 36 selected comparison stars are cross matched with variable stars of JO20, whom stellar variability is not confirmed in the present analysis. Thus, we concluded that the time series data of one night could lead to an over-estimation of short periodic variables via absolute/standard phtotmetry. Since computed value of period for variable star $V_{3}$ of cluster DO14 is found to be nearly equal to time duration of its observation, therefore author did not classify this star.   

JO20 used continuous time series data less than 3.5 hours for each observational night, whereas author has additional time series data of 5.4, 7.6, 7.2 and 5.6 hours as observed on dated 11-12-2013, 20-12-2013, 12-01-2015 and 08-02-2015 respectively. Since period value of variable star $V_2$ of M36 by JO20 is appears to be inaccurate in the view of additional data, therefore it is best to avoid to classifying variable stars, that have approximately same/ slightly longer period than the time duration of observation for a particular night. 

In the case of DO14, author has constructed light curves for selected comparison stars by magnitude transformation of non-standardized field via SSM approach, whereas light curves for selected comparison stars of M36 have been constructed by absolute photometry via SSM approach. Their comparative analysis indicates that absolute photometry is a meaningful  to search for variable stars in the target filed.  SSM approach gave similar trends of pseudo-variability for comparison stars in the field of  both cluster. As a result, this approach confirms that resulting light curves of comparison stars do not depend on the nature of reference catalogue produced by either standard or instrumental magnitudes of stars.

Author concludes that the detection process of short-period variability shorter than a single night has nothing to do with the absolute/standard photometry. Although, the appearance of a unique feature in light curves of potential variable star is confirmed through the differential photometry of similar stars.

 
%
\section{Conclusion}
%

 
 
The present analysis is devoted the detection of short periodic variable stars with periods less than 1 day. In this connection, author did not present any analysis for the classification of potential candidate of variable stars characterized by long periodicity and irregular variability. To find small periodic variables, time series data for DOLIDZE\, 14 and NGC\,1960 have been collected over one and five nights of observations, respectively. In this connection, the stellar light curves for NGC\,1960 and DOLIDZE\, 14 have been extracted from these time series data.\\
By deep investigate of stellar light curves, a total of 18 and 4 variable stars have been identified in the field of view of cluster NGC\,1960 and DOLIDZE\,14 respectively. 
 Four of these 18 variable stars are cross matched with JO20. In addition, other three variable stars of JO20 are not found to be short pulsation stars in present analysis. To perform membership analysis for variable stars, the mean proper motion of NGC~1960 and DOLIDZE~14 in their RA and DEC directions were estimated as ($0.314{\pm}0.026~mass/yr$ and $-2.333{\pm}0.037~mass/yr$) and ($1.050{\pm}0.083~mass/yr$ and $-2.254{\pm}0.093~mass/yr$), respectively.\\
The time duration (1.2 yr) seems good enough to accurately determine the Period of any variable star. As per 10 cases out of 18 for NGC~1960 ($V_{1}$, $V_{3}$, $V_{4}$, $V_{5}$, $V_{8}$, $V_{9}$, $V_{13}$, $V_{14}$, $V_{15}$ and $V_{16}$), most derived periods are 0.25-0.30 days in length, this is the average length of the individual data strings in 2013 and 2015. This makes the period determination very suspicious. Thus, the large gap in between the nights (almost 1 yr) causes additional aliases. As a result, author concludes that it is impossible to determine the true periods, and the physical reason of the light variability based on such scare data sets poorly dispersed over a long period of time.\\
 Due to the observational limitations of CCD camera of 1.04 m telescope at ARIES, a telescope equipped with a very high-capacity CCD camera is needed to carry out the task of searching for a signal of variability in the brighter stars of NGC 1960.
\section{Competing interest}
The authors declare that they have no competing interest.
\section{Consent for publication}
Not applicable.
\section{Ethics approval and Consent to participate}
Not applicable.
\section*{Acknowdege}\label{ss:ack}
GCJ acknowledges the web-portal services of SIMBAD, VIZIER and ESO. Collection and observation of time series data of both clusters (NGC 1960 and DOLIDZE 14) performed by GCJ during 18 September 2012 to 27 April 2015. Director, ARIES gave permission to GCJ to use ARIES Data for research work via letter no. AO/2018/41 on date 12 April 2018.
%

%
%
%
\begin{figure*}
\includegraphics[width=16.0cm]{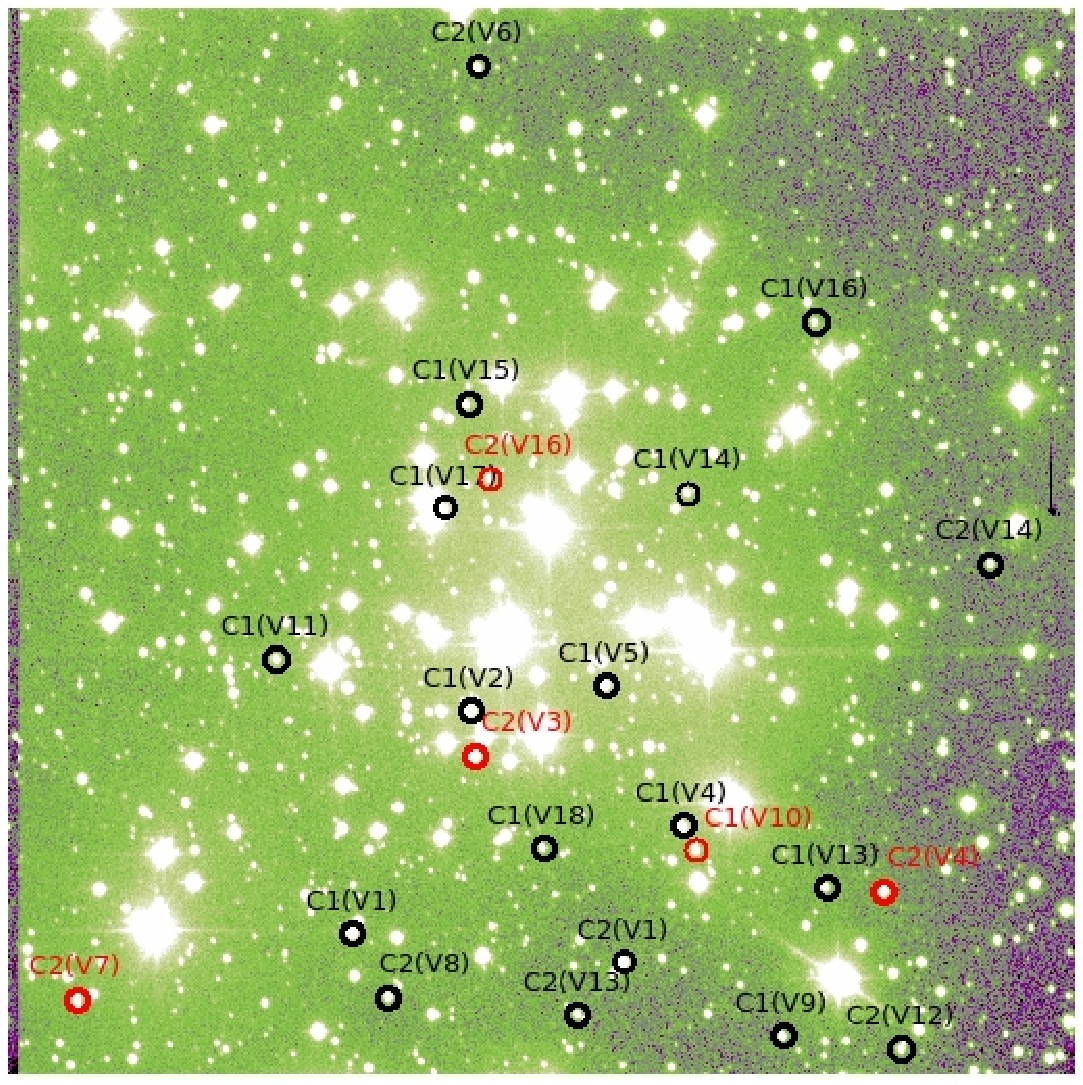}
\caption{The finding chart for selected comparison stars within cluster NGC~1960. Red open circles represent those Long term variable stars, for which variation in brightness is also found as daily basis. Black open circles are depicted stars that have irregular flux variation in their light curves as extracted through absolute photometry.}
\label{fig14}
\end{figure*}

\begin{figure*}
\centerline{\includegraphics[width=16.0cm]{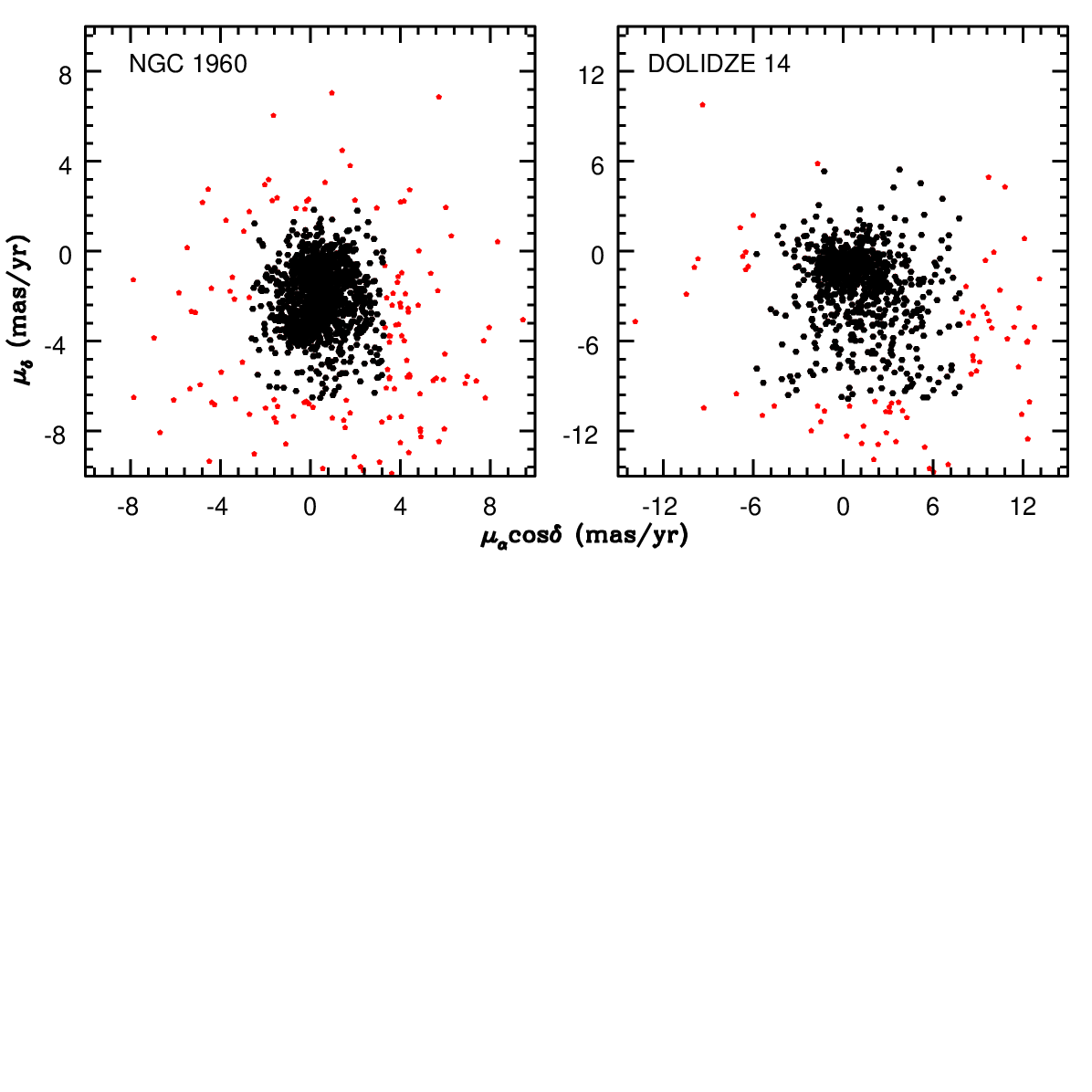}}
\vspace{-7.5cm}
\caption{The distribution of proper motion of stars present in the cluster region. The large points represent those stars which are used to determine the mean proper motion of the cluster.}
\label{Fig15}
\end{figure*}

\begin{figure*}
\centerline{\includegraphics[width=16.0cm]{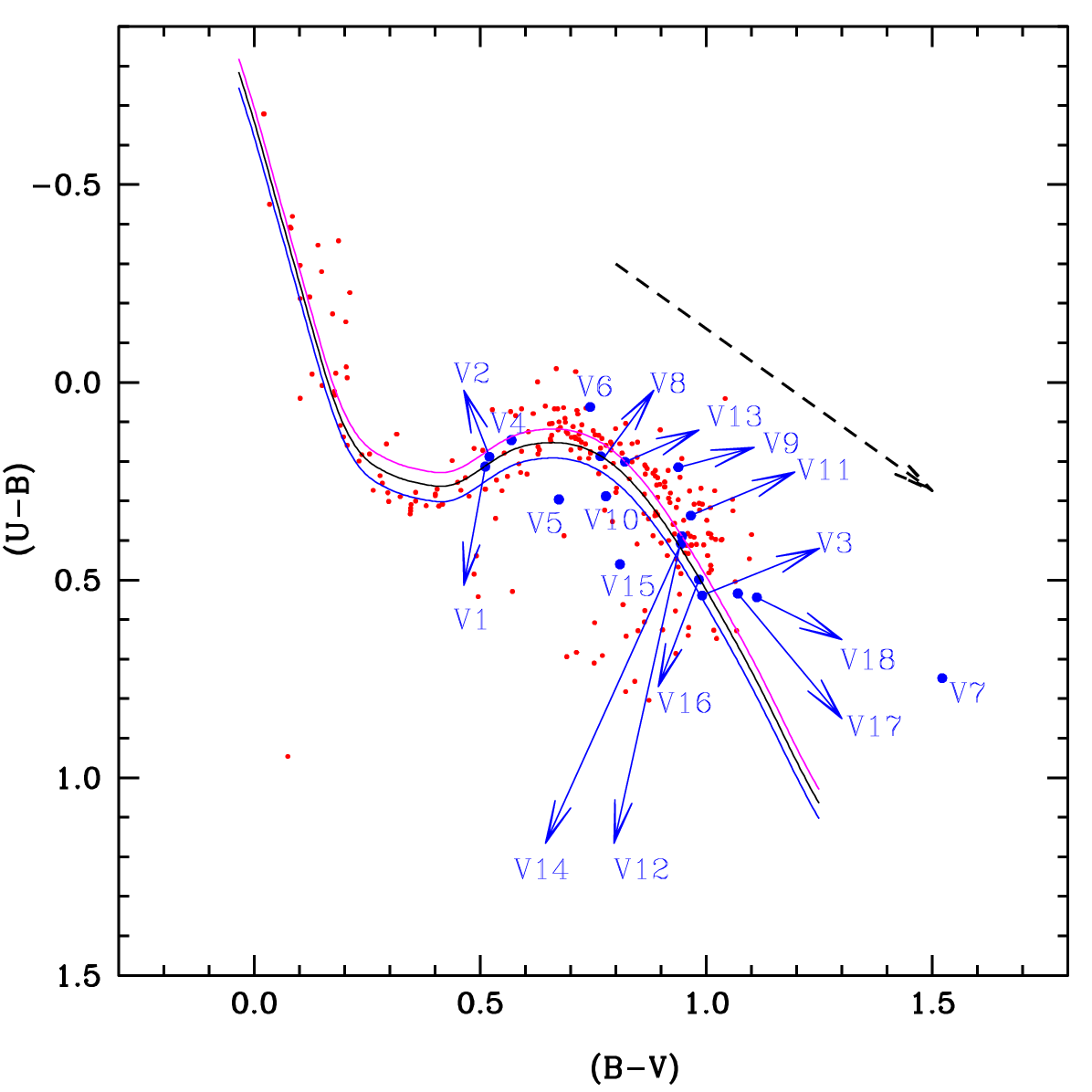}}
\caption{$(B-V)$ vs $(U-B)$ colour-colour diagram for the MPMs in the field of cluster NGC\,1960 as per Joshi {\&} Tyagi (2015). For clarity, variable stars of NGC 1960 are represented by blue dots. The dotted black arrow shows a slope of normal reddening vector $E(U-B)/E(B-V)$=0.72. The solid black line shows the best fit taken in account of 0.23 and 0.17 mag shift in $(B-V)$ and $(U-B)$, respectively while pink and blue lines represent an error of $\pm$0.15 in the reddening vector respectively.}
\label{fig15}
\end{figure*}
\begin{figure*}
\includegraphics[width=16.0cm]{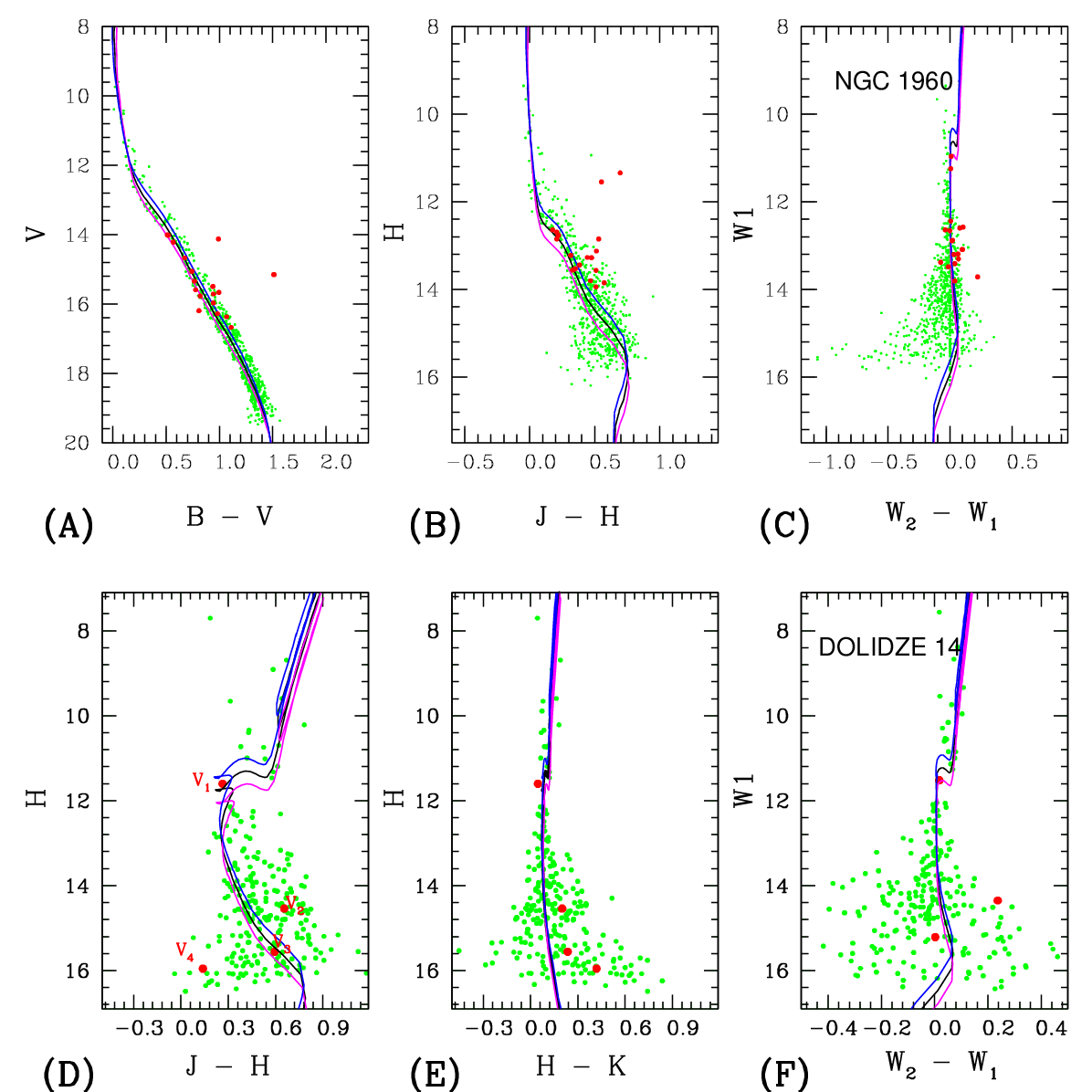}
\caption{The upper and bottom panels of this figure represent CMDs for NGC 1960 and DOLIDZE 14, respectively. The red dots on each panel represent the variable stars as identified by us through the time series photometric data while green dots represent the probable members of the studied clusters as extracted from the work of Joshi {\&} Tyagi (2015) and Joshi {\&} Tyagi (2015b). The black solid lines represent the best fitted theoretical isochrones as given in the previous studies. In the panel {\bf 16 (A)}, there are no variable stars situated near the turn-off region of NGC~1960 due to selected magnitude range ($13.17{\pm}0.30 ~mag$ to $16.61{\pm}0.30~ mag$) in V-band. In the case of no prerequisite criteria of magnitude, detected variable stars should be found simultaneously everywhere including (turn-off region) as depicted for DOLIDZE~14 in the panel {\bf 16 (D)}.}
\label{fig16}
\end{figure*}

\end{document}